\documentclass[twocolumn,secnumarabic,amssymb, nobibnotes, aps, prd, superscriptaddress]{revtex4}

\usepackage[utf8]{inputenc}
\usepackage{graphics,graphicx}
\usepackage{amsmath,amssymb,amsfonts,latexsym,cancel}
\usepackage{bm}

\begin{document}
\title{Equilibrium and stability of charged strange quark stars}

\author{Jos\'e D. V. Arba\~nil}\email{arbanil@ita.br}
\affiliation{Departamento de Física, Instituto Tecnológico de Aeronáutica, Centro Técnico Aeroespacial, 12228-900 São José dos Campos, SP, Brazil.}
\author{M. Malheiro}\email{malheiro@ita.br}
\affiliation{Departamento de Física, Instituto Tecnológico de Aeronáutica, Centro Técnico Aeroespacial, 12228-900 São José dos Campos, SP, Brazil.}

\begin{abstract} 
The hydrostatic equilibrium and the stability against radial perturbation of charged strange quark stars composed of a charged perfect fluid are studied. For this purpose, it is considered that the perfect fluid follows the MIT bag model equation of state and the radial charge distribution follows a power-law. The hydrostatic equilibrium and the stability of charged strange stars are investigated through the numerical solutions of the Tolman-Oppenheimer-Volkoff equation and the Chandrasekhar's pulsation equation, being these equations modified from their original form to include the electrical charge. In order to appreciably affect the stellar structure, it is found that the total charge should be of order $10^{20}[\rm C]$, implying an electric field of around $10^{22}[\rm V/m]$. We found the electric charge that produces considerable effect on the structure and stability of the object is close to the star's surface. We obtain that for a range of central energy density the stability of the star decreases with the increment of the total charge and for a range of total mass the electric charge helps to grow the stability of the stars under study. We show that the central energy density used to reach the maximum mass value is the same used to determine the zero eigenfrequency of the fundamental mode when the total charge is fixed, thus indicating that the maximum mass point marks the onset of instability. In other words, when fixing the total charge, the conditions $\frac{dM}{d\rho_c}>0$ and $\frac{dM}{d\rho_c}<0$ are necessary and sufficient to determine the stable and unstable equilibrium configurations regions against radial oscillations. We also consider another charge distribution, charge density proportional to the energy density, and show that our results do not depend on this choice and the conditions used to determine regions made of the stable and unstable charged equilibrium configurations are maintained.
\end{abstract}
 \maketitle

\section {Introduction}
\label{sec-introd}

\subsection{Hydrostatic equilibrium of charged strange stars}

The theoretical possibility that strange quark matter (strange matter, for short) can be the ground state for the strong interaction was proposed, independently, by Bodmer \cite{bodmer1971} and Witten \cite{witten1984}. This matter is usually thought as a liquid with equal numbers of unconfined up, down and strange quarks. Some authors have studied how this strange material could manifest in nature. Among the possibilities, the study of stars composed by strange matter has gained attention. Nowadays these spherical objects are known as strange quark stars or, simply, strange stars.

Whether stars made of strange matter exist, these could be a structure made of approximately equal number of up, down and strange quarks together with smaller number of electrons required to provide electrical charge neutrality. In a strange star the electrons are distributed on its surface, forming an electron surface width of several hundred fermis. The strong electric field generated in this region is on the order of $10^{20}[\rm V/m]$ (see, e.g., \cite{alcock1986,alcock1988}). This strong surface electric field value may be exceeded if strange matter forms a color superconductor, as expected for such matter \cite{alford2008}.

Strange quark stars with a strong electric field can be modeled solving the  Maxwell-Einstein field equations. Through the numerical solution of the Tolman-Oppenheimer-Volkoff (TOV) equation for the charged case \cite{bekenstein}, also known as the hydrostatic equilibrium equation, Negreiros and collaborators \cite{negreiros2009,negreiros2011} modeled these objects considering spheres composed of strange matter that follows the MIT bag model equation of state (EoS) and a Gaussian distribution of the electric charge on the surface of the star. Through this model, the authors estimated that the electric charge that causes significant impact on the structure of the strange stars produces a surface electric field of the order $E\sim10^{22}[\rm V/m]$. This strong electric field is found in stars where the electrical energy density is not negligible compared to the radial pressure. 

The studies developed in \cite{negreiros2009,negreiros2011} form part of a quantity of works where the influence of the electric
charge on the structure of spherically symmetric static objects is analyzed. Within this bulk of works, we found studies developed considering polytropic stars with a charge density proportional to the energy density \cite{raymalheirolemoszanchin,siffert,alz-poli-qbh} and with a charge  density proportional to the rest mass density \cite{ghezzi2005}. We also found models where incompressible stars (stars with a constant energy density) are considered with an electric charge distribution $q(r)$ following a power-law of the form $q(r)=Q (r/R)^n$. In \cite{defelice_yu,defelice_siming},  $n\geq3$ is taken in order to prevent the divergence of the charge density at the origin, with parameters $Q$ and $R$ representing the total charge and the star's radius. Anninos and Rothman also investigated the structure of charge incompressible stars in \cite{annroth} considering a complex distribution of electric charge dependent on the radial coordinate. For others examples review  \cite{alz-incom-qbh} and references cited therein.

\subsection{Stability of charged strange stars}

Even though the first work on the stability of electrically charged stars was developed  in the Maxwell-Einstein context in the seventies, we have not found studies on electrically charged strange stars. In the literature, we find that the first works on stability of charged stars was investigated by Stettner \cite{stettner1973}. He analyzed the stability problem of a homogeneous distribution of matter containing a constant surface charge. Stettner found that stars with a constant energy density and with a little surface charge are more stable than uncharged incompressible stars. Some years latter, Glazer generalized the Chandrasekhar's equation of pulsation to include the electric charge \cite{glazer1976,glazer1979} and studied the stability of Bonnor's charged dust stars and charged  incompressible stars. Glazer presented in \cite{glazer1976} that the Bonnor stars are dynamically unstable and in \cite{glazer1979} that the electric charge increases the stability of an incompressible star. An analysis of the stability of incompressible charged stars was also performed in \cite{defelice_siming,annroth} where it was concluded that for a range of parameters these configurations can be stable. The stability  against radial oscillations for a charged star with a realistic EoS was investigated by Brillante and Mishustin \cite{brillante2014}. They studied hybrid stars with a small net electric charge. The authors showed that, for a fixed value of the central baryon density, an increment of the electric charge leads to lower values of eigenfrequencies.

\subsection{This work}

We study how the equilibrium structure and stability of strange stars are affected by electric charges. With that purpose we use the MIT bag model EoS and a distribution of electric charge of the form $q(r)=Q(r/R)^3$ ($\equiv\beta r^3$, being $\beta=Q/R^3$). The results obtained assuming that the charge distribution follows this power-law are supplement with others obtained assuming that the charge density is proportional to the energy density of the form $\rho_e=\alpha\rho$ (being $\alpha$ a dimensionless proportionality constant).

The paper is structured as follows. In Sec.~\ref{eos1} we describe the EoS used and the profile of the electric charge considered. In Sec.~\ref{tov_roe} we show the hydrostatic equilibrium equations  and radial oscillation equations, both  modified to include a net electric charge, as well as the numerical method used to numerically solve these equations. In Secs.~\ref{resuls} and ~\ref{E_S_Q} the studies of equilibrium and stability of charged strange quark stars with fixed $\beta$ and with fixed total charge $Q$ are presented, respectively. In Sec.~\ref{supplement} we supplement the results obtained in Secs.~\ref{resuls} and ~\ref{E_S_Q} using a different charge distribution, specifically we use the charge density proportional to the energy density $\rho_e=\alpha\rho$. Finally, in Sec.~\ref{conclusion} we present our conclusions. Throughout the present article we use the units $c=1=G$.

\section{Equation of state and distribution of the electric charge}\label{eos1}

We use the strange quark matter equation of state of the MIT bag model, i.e.,
\begin{equation}\label{eos}
p=\left(\rho-4{\cal B}\right)/3.
\end{equation}
The parameters $p$ and $\rho$ represent respectively the pressure and the energy density of the fluid, and ${\cal B}$ the bag constant. We consider for the bag constant ${\cal B}=60\,[\rm MeV/fm^3]$ because, as stated in \cite{witten1984,haensel}, with this bag constant the parameters of the maximum mass configuration for strange stars are similar to those for realistic neutron star made of baryonic matter. 

Since our purpose is to investigate the effects of the electric charge on the structure of strange stars, we also have to define the electric charge distribution. We assume that the electric charge follows a function of the radial coordinate of the form:
\begin{equation}\label{charge_distribution}
q=Q\left(\frac{r}{R}\right)^{3}\equiv\beta r^{3},
\end{equation}
with $Q$ and $R$ being the total charge and the total radius, respectively. As  show in Eq.~\eqref{charge_distribution}, $\beta$ is a constant that is related to the total charge and the total radius of the star of the form $\beta=Q/R^3$. The ansatz considered for the electrical charge distribution is the simplest case of electric charge distribution used in \cite{defelice_yu,defelice_siming}.

\section{Stellar structure equations and Radial oscillations equations}\label{tov_roe}

\subsection{Stellar structure equations}

The unperturbed line element used to describe a spherical charged object, in Schwarzschild-like coordinates, is given by:
\begin{equation}\label{metric1}
ds^2=-e^{\nu}dt^2+e^{\lambda}dr^2+r^2\left(d\theta^{2}+\sin^2\theta d\phi^2\right),
\end{equation}
where the exponents $\nu=\nu(r)$ and $\lambda=\lambda(r)$ are functions of the radial coordinate $r$ alone. As mentioned before, the matter contained in the charged sphere is described by the charged perfect fluid energy-momentum tensor. In the considered spacetime and energy-momentum tensor, the equations that govern the hydrostatic stellar structure of a charged sphere are:
\begin{eqnarray}
&&\hspace{-0.2cm}\frac{dq}{dr}=4\pi\rho_{e}r^{2}e^{\lambda/2},\label{qo}\\
&&\hspace{-0.2cm}\frac{dm}{dr}=4\pi\rho r^{2}+\frac{q}{r}\frac{dq}{dr},\label{mo}\\
&&\hspace{-0.2cm}\frac{dp}{dr}=-\left(p+\rho\right)\left(4\pi rp+\frac{m}{r^2}-\frac{q^2}{r^3}\right)e^{\lambda}+\frac{q}{4\pi r^{4}}\frac{dq}{dr},\label{tov}\\
&&\hspace{-0.2cm}\frac{d\nu}{dr}=-\frac{2}{(p+\rho)}\left(\frac{dp}{dr}-\frac{q}{4\pi r^{4}}\frac{dq}{dr}\right),\label{nuo}
\end{eqnarray}
with the potential metric $e^{\lambda}$ of the form:
\begin{equation}
e^{-\lambda}=1-\frac{2m}{r}+\frac{q^2}{r^2}.
\end{equation}
As usual, the parameters $q$ and $m$ represent the charge and mass within the radius $r$ and $\rho_e$ the electric charge density. 
Eq.~\eqref{tov} is the Tolman-Oppenheimer-Volkoff equation \cite{tolman,oppievolkoff}, also known as the hydrostatic equilibrium equation, with the inclusion of the electric charge (see, e.g., \cite{bekenstein,negreiros2009,negreiros2011,raymalheirolemoszanchin,siffert,alz-poli-qbh,alz-incom-qbh}). 

The stellar structure equations, Eqs.~\eqref{qo}-\eqref{nuo}, are integrated from the center toward the surface of the star. The integration of these equations starts with the values at the center:
\begin{equation}\label{condition_center}
m(0)=0,\;\; q(0)=0,\;\; \rho(0)=\rho_c,\;\; {\rm and}\;\; \nu(0)=\nu_c.
\end{equation}
The surface of the star is determined by $p(R)=0$. At this point the interior solution is smoothly connected to the vacuum exterior 
Reissner-Nordstr\"om metric. This means that, at the surface of the charged star, the inner and outer potential functions are related through the equality:
\begin{eqnarray}\label{condition}
e^{\nu(R)}=e^{-\lambda(R)}=1-\frac{2M}{R}+\frac{Q^2}{R^2},
\end{eqnarray}
with $M$ being the total mass of the star. The boundary condition of the functions $\nu$ and $\lambda$ at the surface of the star are determined by relation \eqref{condition}.

\subsection{Radial oscillations equations}

The equations governing infinitesimal radial oscillations are determined  by perturbing the fluid and spacetime variables. This is done in a manner that the spherical symmetry of the background object is maintained. The perturbations are inserted in the field equations and in the energy-momentum tensor conservation while maintaining only first order terms.

The equation for infinitesimal radial oscillations of an uncharged spherical object was obtained by Chandrasekhar \cite{Chandrasekhar1964-a,Chandrasekhar1964-b}. This equation constitutes a Sturm-Liouville eigenvalue problem, its solution furnishes the eigenvalues and eigenfunction of the radial perturbation. Aiming to obtain a more advantageous equation for numerical applications, the Chandrasekhar's pulsation equation can be placed in a different form. The pulsation equation can be derived into two first-order equations for the quantities $\Delta r/r$ and $\Delta p$ \cite{gondek1997}, with $\Delta r$ and $\Delta p$ being respectively the relative radial displacement and the Lagrangian perturbation of pressure (review also \cite{lugones2010}). 

The infinitesimal radial pulsation was also investigated considering the effect of the electric charge in stars. This is done by considering the effect of the electric charge in the energy-momentum tensor, and thus extending the Chandrasekhar's equation for radial pulsation  for the case of a charged star. The equation for radial oscillations of charged objects has been presented in a more appropriate form for numerical solutions. Brillante and Mishustin \cite{brillante2014} derived the radial oscillation equation for the charged case into two first-order equations for the quantities $\Delta r/r$ and $\Delta p$. This set of equations are a generalized form of the first-order set of equations presented in \cite{gondek1997}. The system of equations for charged stars is, $\xi=\Delta r/r$:
\begin{eqnarray}
&&\frac{d\xi}{dr}=\frac{\xi}{2}\frac{d\nu}{dr}
-\frac{1}{r}\left(3\xi+\frac{\Delta p}{p\Gamma}\right),\label{ro1}\\
&&\frac{d\Delta p}{dr}=e^{\lambda-\nu}(p+\rho)\omega^2\xi r+\left(\frac{d\nu}{dr}\right)^2\frac{(p+\rho)\xi r}{4}\nonumber\\
&&-4\xi\left(\frac{dp}{dr}\right)-8\pi(p+\rho)\xi re^{\lambda}\left(p+\frac{q^2}{8\pi r^4}\right)\nonumber\\
&&-\left(\frac{1}{2}\frac{d\nu}{dr}+4\pi re^{\lambda}(p+\rho)\right)\Delta p,\label{ro2}
\end{eqnarray}
with $\Gamma=\left(1+\frac{\rho}{p}\right)\left(\frac{dp}{d\rho}\right)$, $\omega$ being the eigenfrequency, and the quantities $\xi, \Delta p$ are assumed to have time dependence $e^{i\omega t}$. To solve Eqs.~\eqref{ro1} and \eqref{ro2} two boundary conditions are necessary. In order to have a regular solution in the center of the star, it is required that the coefficient in Eq.~\eqref{ro1} vanishes for $r\rightarrow0$:
\begin{equation}\label{bc_oscillation}
(\Delta p)_{\rm center}=-3(\xi\Gamma p)_{\rm center}.
\end{equation}
For normalized eigenfunctions we have $\xi(r=0)=1$ at the center of the star. On the surface of the star $r=R$, $p\rightarrow 0$, implying that:
\begin{equation}\label{bc_oscillation1}
(\Delta p)_{\rm surface}=0.
\end{equation}

\subsection{Numerical method}\label{numerical_method}

{\bf Hydrostatic equilibrium of charged strange stars:} Once given the EoS and electric charge distribution, Eqs. \eqref{eos} and \eqref{charge_distribution} respectively, we numerically solve Eqs.~\eqref{qo}-\eqref{nuo} together with the boundary conditions \eqref{condition_center} and \eqref{condition} for different values of $\beta$ and $\rho_c$, by using the Runge-Kutta method of fourth order implemented with the shooting method. 

The numerical solutions start with the integrations of Eqs.~\eqref{qo}-\eqref{tov}. These equations are integrated from the center to the surface of the object through the Runge-Kutta method for a given $\rho_c$ and $\beta$. 

After obtaining the coefficients $p$, $\rho$, $\rho_e$, $q$, $m$ and $\lambda$ for a given $\rho_c$ and $\beta$, we use the shooting method to solve Eq.~\eqref{nuo} because its boundary condition Eq.~\eqref{condition} is at the surface of the star. In order to integrate numerically Eq.~\eqref{nuo} we consider a proof value $\nu_c$ in the center of the star. Eq.~\eqref{nuo} is integrated from the center toward the surface of the object. Whether after the integration the boundary condition \eqref{condition} is not satisfied, we correct the value of $\nu_c$. This process is repeated until the equality \eqref{condition} is satisfied. \\

{\bf Radial oscillations of charged strange stars:} After obtaining the coefficients of the pulsation equation for a given $\rho_c$ and $\beta$, the pulsation equation \eqref{ro1} and \eqref{ro2} and the boundary conditions \eqref{bc_oscillation} and \eqref{bc_oscillation1} were solved by a shooting method using Runge-Kutta integration. The integrations of pulsation equations are developed from the origin to the surface of the star. We start this process considering a trial value for $\omega^2$. Whether after each integration the condition at the surface of the star \eqref{bc_oscillation1} is not satisfied, the value of $\omega^2$ is corrected. The process is repeated until condition \eqref{condition} is satisfied. The values of $\omega^2$ for which the boundary condition is satisfied are called eigenvalues of the pulsation equation and $\omega$ of eigenfrequencies.

\section{Equilibrium and stability of charged strange quark stars with fixed $\beta$}\label{resuls}

\subsection{Equilibrium of charged strange quark stars with fixed $\beta$}\label{equi_beta}

\begin{figure}[!h]
\includegraphics[scale=0.29]{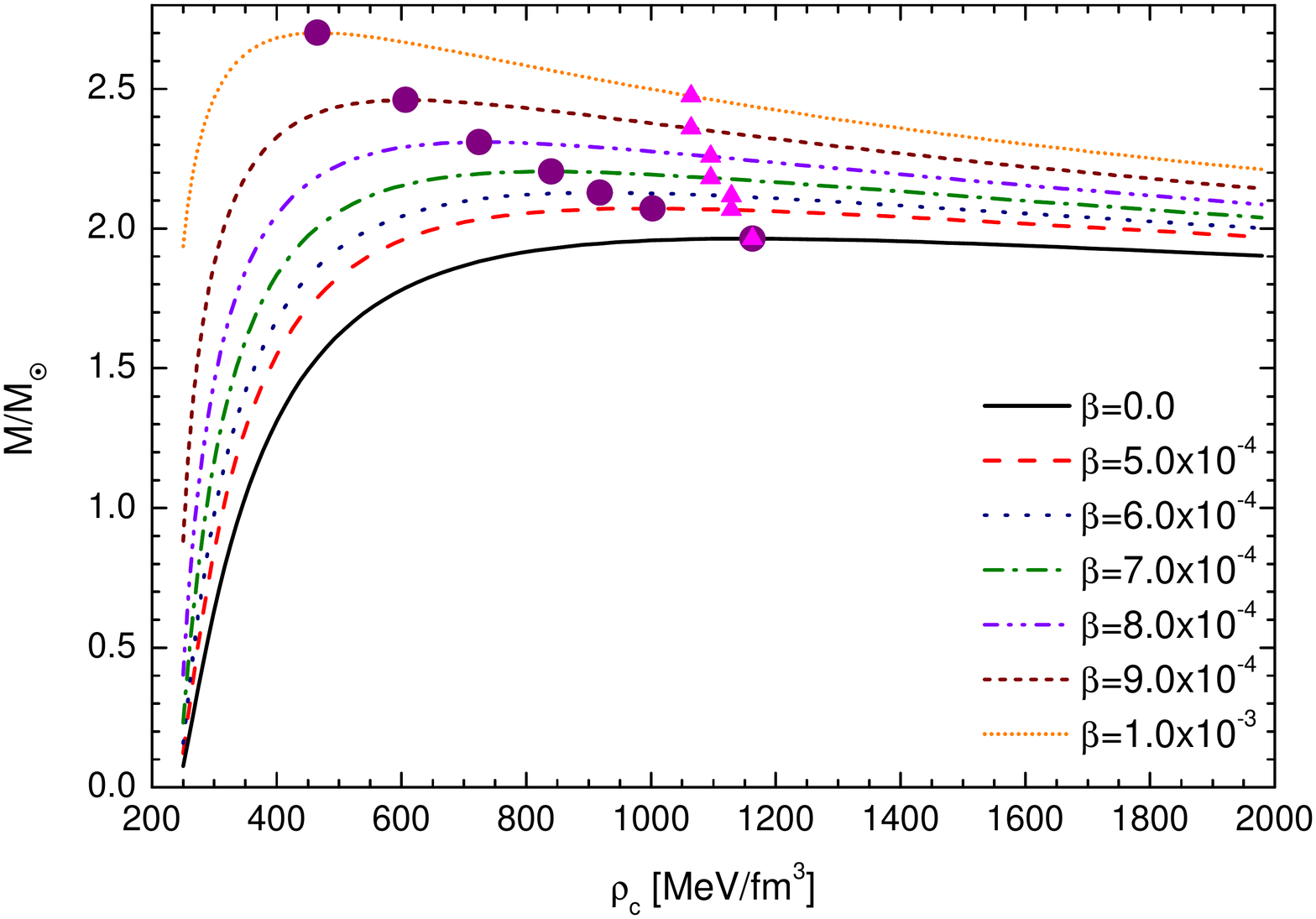}
\vspace*{-.7cm}
\caption{Mass of the star, normalized in solar masses $M_{\odot}$, versus the central energy density $\rho_c$ for some values of $\beta$. The units used for $\beta$ are $[M_{\odot}/{\rm km}^3]$. The full circles represent the maximum mass points and the full triangles the places where the zero eigenfrequencies of the fundamental mode are found.}
\label{rho_M}
\end{figure}

Fig.~\ref{rho_M} shows the stellar mass, in solar masses $M_{\odot}$, against the central energy density $\rho_c$ for different values of the constant $\beta$. We use central energy densities in the range $250\leq\rho_c\leq2000\,[{\rm MeV/fm^3}]$. The full circles and the full triangles over the curves indicate the points where the maximum mass values and the zero eigenfrequencies of the fundamental mode are found, respectively. In all cases, we observe that the mass of the star grows with the increment of the central energy density until it reaches a maximum mass in $\rho_c=\rho_{c}^{*}$. For central energy densities larger than $\rho_{c}^{*}$ the mass decreases with the growth of $\rho_c$. 

For the uncharged case ($\beta=0$) we observe that the central energy density used to reach the maximum mass value coincides with the value of $\rho_c$ considered to determine the eigenfrequency $\omega=0$. This means that the maximum mass point separates the stable equilibrium configuration from the unstable one. The equilibrium configurations lying in the region where $\frac{dM}{d\rho_c}>0$ are always stable configurations, in turn, the configurations that are in the region where $\frac{dM}{d\rho_c}<0$ are always unstable. I.e., the inequalities $\frac{dM}{d\rho_c}>0$ and $\frac{dM}{d\rho_c}<0$ are necessary and sufficient conditions to recognize regions made of stable and unstable equilibrium configurations, respectively. On the other hand, for the charged cases ($\beta\neq0$) we note that the central energy density used to determine the maximum mass point ($\rho_{c}^{*}$) is not the same $\rho_c$ used to find $\omega=0$. We obtain that the zero eigenfrequency is found in a central energy density $\rho_c>\rho_{c}^{*}$. This indicates that the relation $\frac{dM}{d\rho_c}>0$ is only a necessary condition to recognize regions of stable configurations. 
                                          
\begin{figure}[!h]
\includegraphics[scale=0.29]{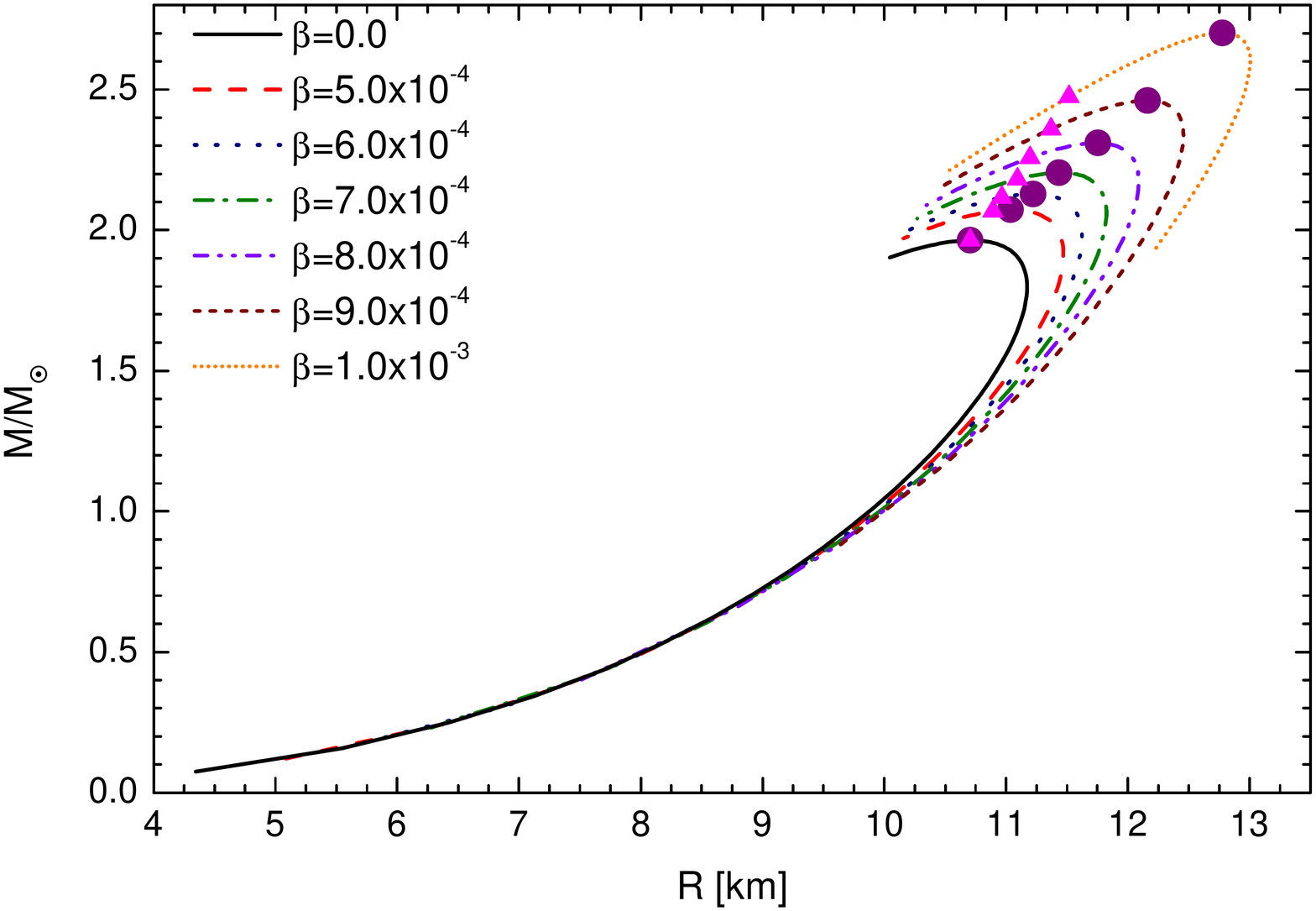}
\vspace*{-.7cm}
\caption{Radius against the mass of the star, $M/M_{\odot}$, for different values of $\beta$, given in units of $[M_{\odot}/{\rm km}^3]$. The full circles and the full triangles represent respectively the maximum mass points and the places where the zero eigenfrequency of the fundamental mode are found.}
\label{R_M}
\end{figure}
\begin{figure}[!h]
\includegraphics[scale=0.29]{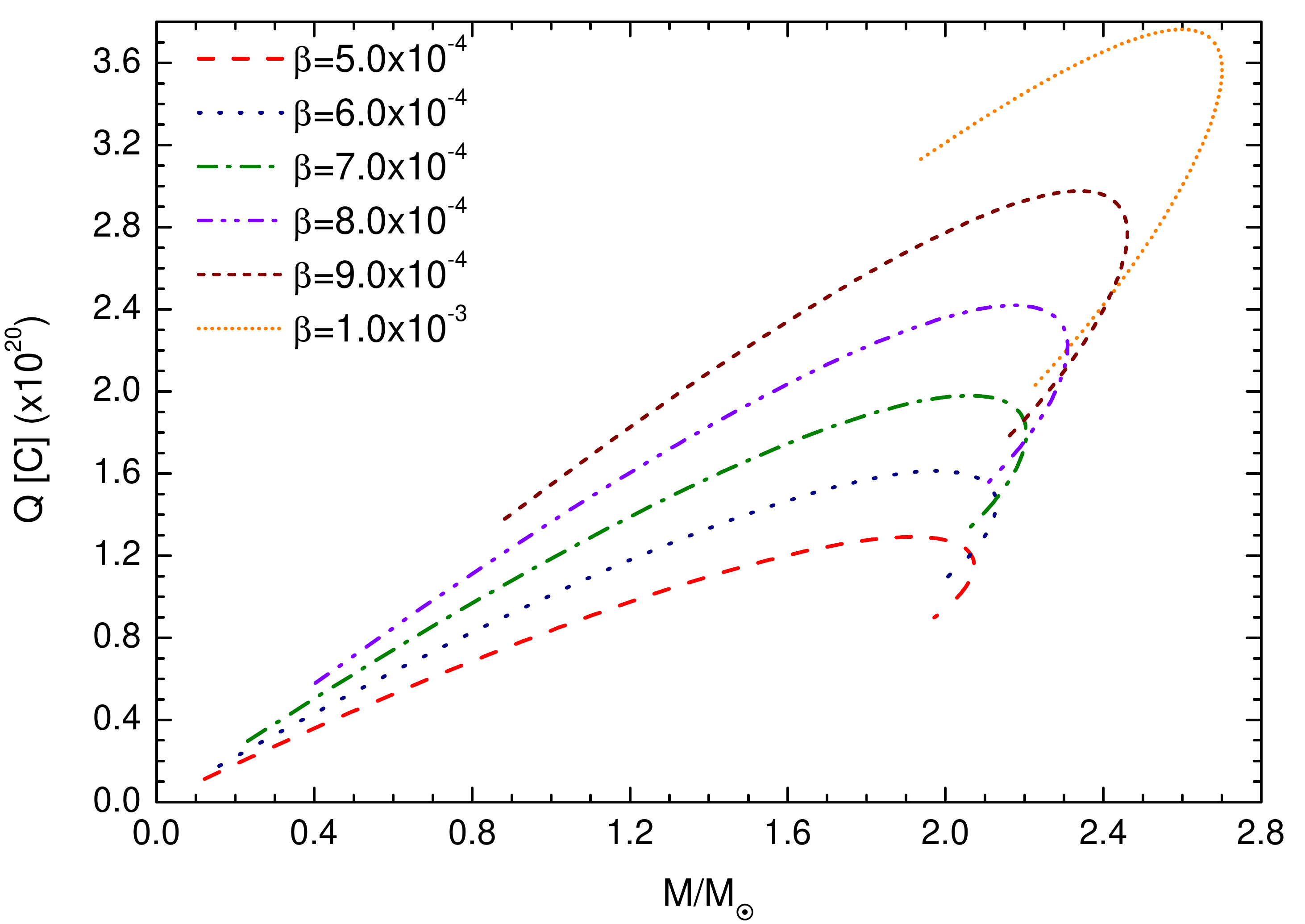}
\vspace*{-.7cm}
\caption{The total charge against the total mass of the star $M/M_{\odot}$ for few values of $\beta$. The units for the constant $\beta$ are $[M_{\odot}/{\rm km}^3]$.}
\label{Q_M}
\end{figure}

In Fig.~\ref{R_M} we present the mass-radius relation for different values of $\beta$. The circles and triangles indicate where the maximum mass points and the zero eigenfrequencies are found. The behavior of the curves shown are characteristic of strange quark stars. It is clear that the masses and radii of the stars change with the increment of $\beta$. For the uncharged case, the point of maximum mass coincides with the point where the zero eigenfrequencies is found. However, as in Fig.~\ref{rho_M}, for $\beta\neq0$ the point of maximum mass does not coincide with the point where $\omega=0$ is determined. Analyzing the maximum mass and its respective radius found in each curve we determine that these values could change from $2\%$ to $30\%$. The growth of the mass and radius with $\beta$ can be explained observing that the electric charge increases with $\beta$ (see Fig.~\ref{Q_M}), so we understand that the electric charge acts as an effective pressure helping the radial pressure to support a more massive star avoiding the collapse.

The total charge as a function of the total mass of the star is plotted in Fig.~\ref{Q_M} for a few values of $\beta$. Note that the total charge grows with the total mass until it reaches the point of maximum charge value. After this point, the charge decreases with the increment of mass. When the maximum mass point is reached, the curves inflect to the left for the charge to begin to decrease with the decay of the electric charge. From this we understand that the value of the maximum charge and the value of the maximum mass do not form part of the same equilibrium configuration. The value of the maximum charge is reached before finding the value of the maximum mass.

It is worth mentioning that in Fig.~\ref{Q_M} it is observed that in some points of the curves we found the limit $Q\approx (M/M_{\odot})\times10^{20}[\rm C]$, using the relation $1\,M_{\odot}=1.7114\times10^{20}\,[\rm C]$ we have $Q\approx M$, this is the same maximum charge limit for the Reissner-Nordstr\"om black hole. 

\begin{figure}[!h]
\includegraphics[scale=0.29]{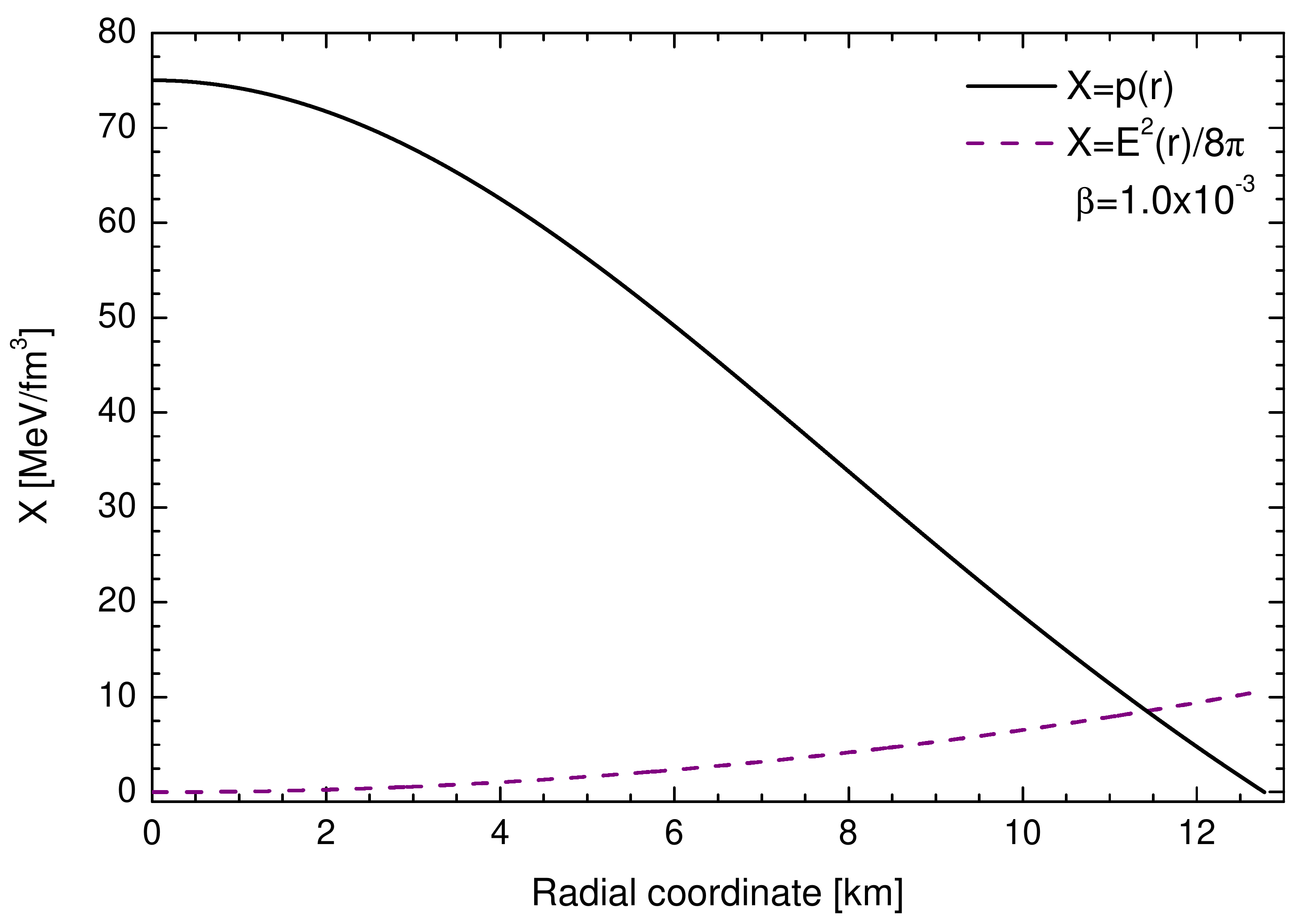}
\vspace*{-.7cm}
\caption{The radial pressure $p(r)$ and the electrical energy density $E^2(r)/8\pi$ as a function of the radial coordinate for a static equilibrium configuration with maximum mass found in $\beta=1.0\times10^{-3}\,[\rm M_{\odot}/km^3]$.}
\label{rc_X}
\end{figure}

\begin{table}[h] \label{table}
\begin{tabular}{ccccc}
\hline\hline
$\beta\,[M_{\odot}/{\rm km^3}]$ & $M/M_{\odot}$ & $R\,[\rm km]$ & $Q\,[\rm C]$ & $E\,[\rm V/m]$\\ \hline
$5.0\times10^{-4}$ & $2.072$ & $11.04$ & $1.151\times10^{20}$ & $8.500\times10^{21}$\\
$6.0\times10^{-4}$ & $2.129$ & $11.22$ & $1.452\times10^{20}$ & $1.037\times10^{22}$\\
$7.0\times10^{-4}$ & $2.205$ & $11.44$ & $1.791\times10^{20}$ & $1.233\times10^{22}$\\
$8.0\times10^{-4}$ & $2.309$ & $11.75$ & $2.223\times10^{20}$ & $1.448\times10^{22}$\\
$9.0\times10^{-4}$ & $2.461$ & $12.16$ & $2.772\times10^{20}$ & $1.686\times10^{22}$\\
$1.0\times10^{-3}$ & $2.701$ & $12.78$ & $3.570\times10^{20}$ & $1.968\times10^{22}$\\
\hline\hline
\end{tabular}
\caption{The constant $\beta$ and the maximum masses with their respective radii, charges and electric fields of the stars.}
\end{table}
The values used for $\beta$ and the maximum masses values with their respective radii, charges, and electric field of the charged spherical objects, are shown in Table~I. Note that some maximum masses and their respective radii and charges are similar to those found in Table I of \cite{negreiros2009}. This indicates that, as well as the Gaussian charge distribution considered in \cite{negreiros2009}, the radial distribution of the electric charge produces considerable effects only near the surface of the strange star. Near the surface of the object, the electrical energy density $E^2(r)/8\pi (\equiv q^{2}(r)/8\pi r^{4})$ is not negligible compared to the radial pressure $p(r)$  (see Fig.~\ref{rc_X}). 

Fig.~\ref{rc_X} shows the behavior of the radial pressure and electrical energy density with the radial coordinate, for a static equilibrium configuration with maximum mass determined with $\beta=1.0\times10^{-3}\,[\rm M_{\odot}/km^3]$.

\subsection{Stability of charged strange quark stars with fixed $\beta$}

\begin{figure}[!h]
\includegraphics[scale=0.29]{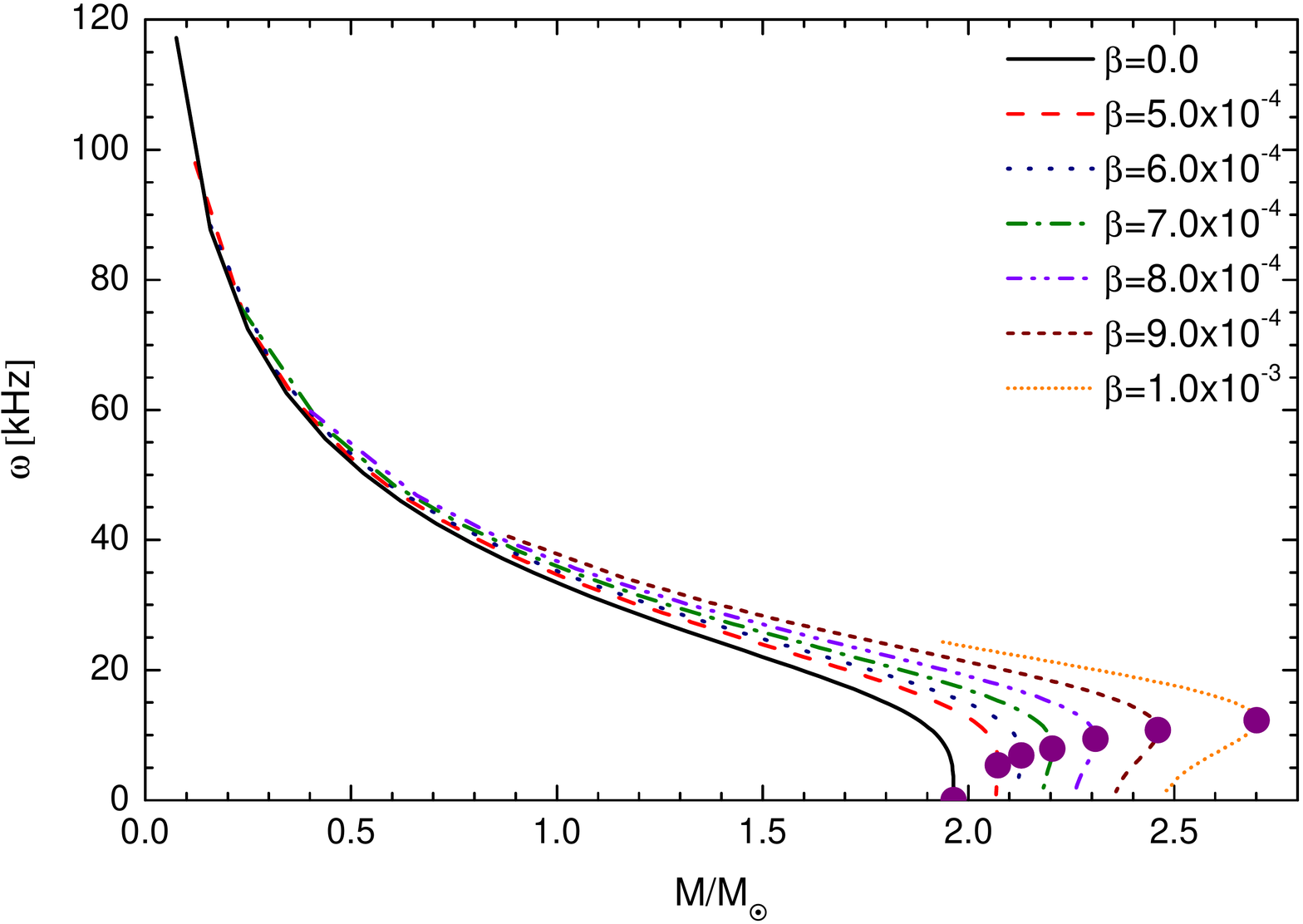}
\vspace*{-.7cm}
\caption{The eigenfrequency of the fundamental mode as a function of the stellar mass of strange charged quark stars, for some different values of $\beta$. In each curve, the full circles represent the maximum mass points. The constant $\beta$ has units of $[M_{\odot}/{\rm km}^3]$. }
\label{omega_M}
\end{figure}

The eigenfrequency of the fundamental mode against the mass, normalized to the Sun's mass, for different values of $\beta$ is shown in Fig.~\ref{omega_M}. The  full circles in purple indicate the points of maximum masses. It is worth mentioning that the curve found for the uncharged case ($\beta=0$) is similar to the curve found in the study of radial oscillations of strange quark stars in \cite{gondek1999} (specifically, Fig.~$7$). For the zero-charge, we have that the value of $\omega$ decreases monotonically with the growth of the mass of the star. However, for the charged case we determine a different behavior of $\omega$ with $M/M_{\odot}$. We note that the eigenfrequency decays with the increment of the mass until the maximum mass value is reached. After this point, $\omega$ begins to decrease with the diminution of the mass until to attain $\omega=0$ thus showing that the maximum mass point does not match the point of zero eigenfrequency. We also note that for some mass parameters, the increment of $\beta$ allows the stars to become more stable against radial perturbations.

\section{Equilibrium and stability of charged strange quark stars with fixed total charge}\label{E_S_Q}

\subsection{Construction of the curves for the study of equilibrium and stability of strange quark stars with fixed total charge}

\begin{figure}[!ht]
\centering
\includegraphics[scale=0.29]{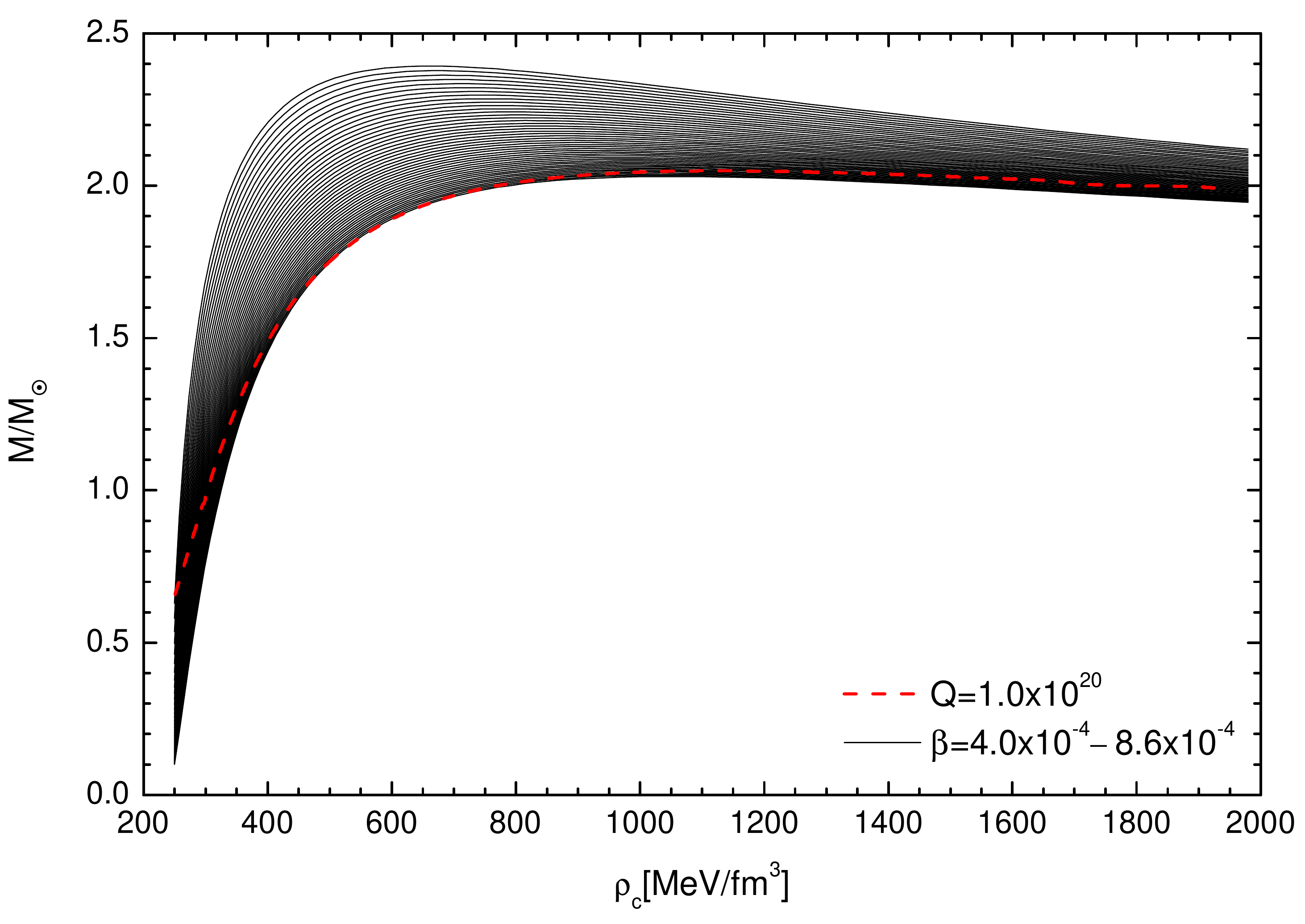}
\centering
\includegraphics[scale=0.29]{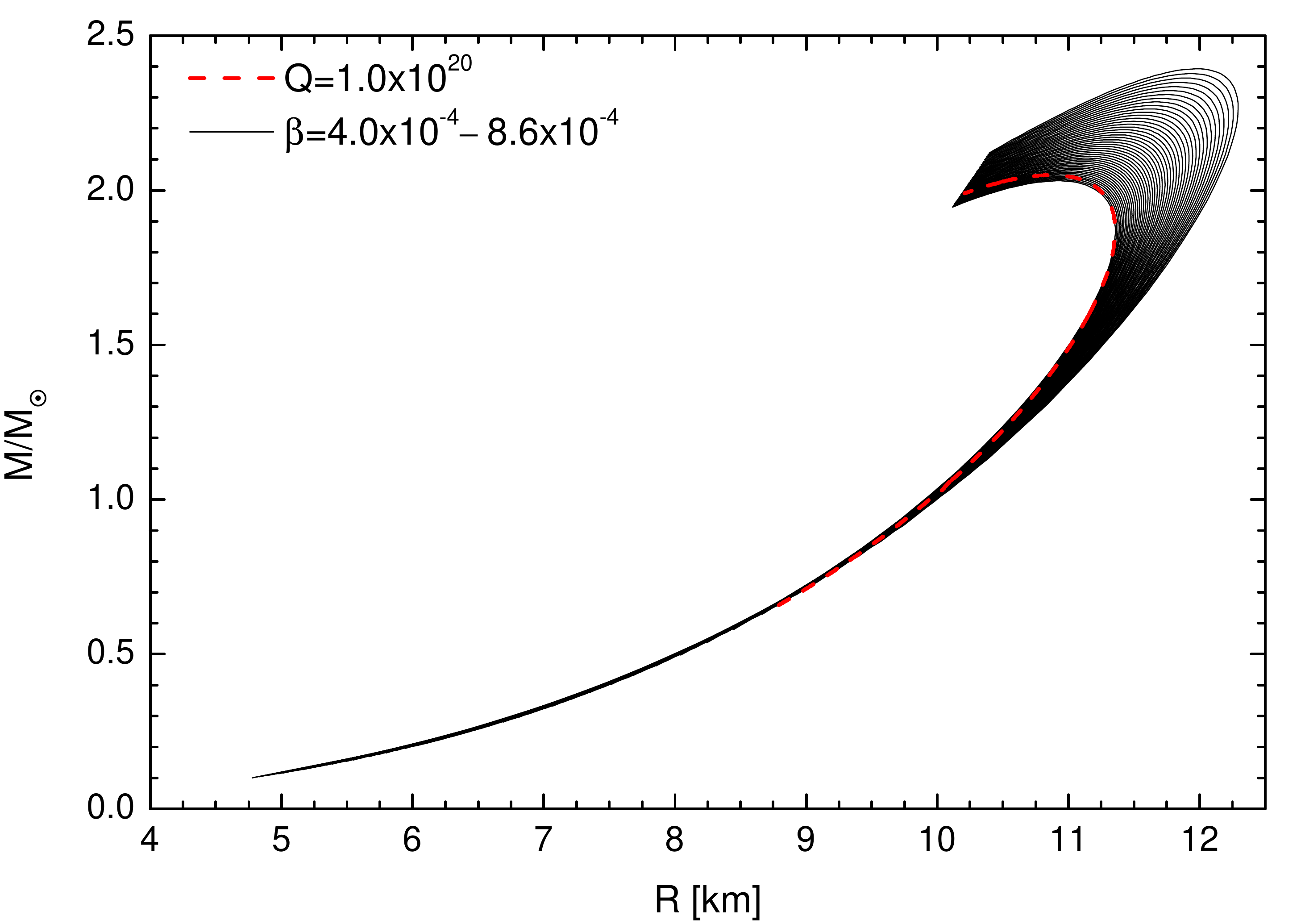}
\centering
\includegraphics[scale=0.29]{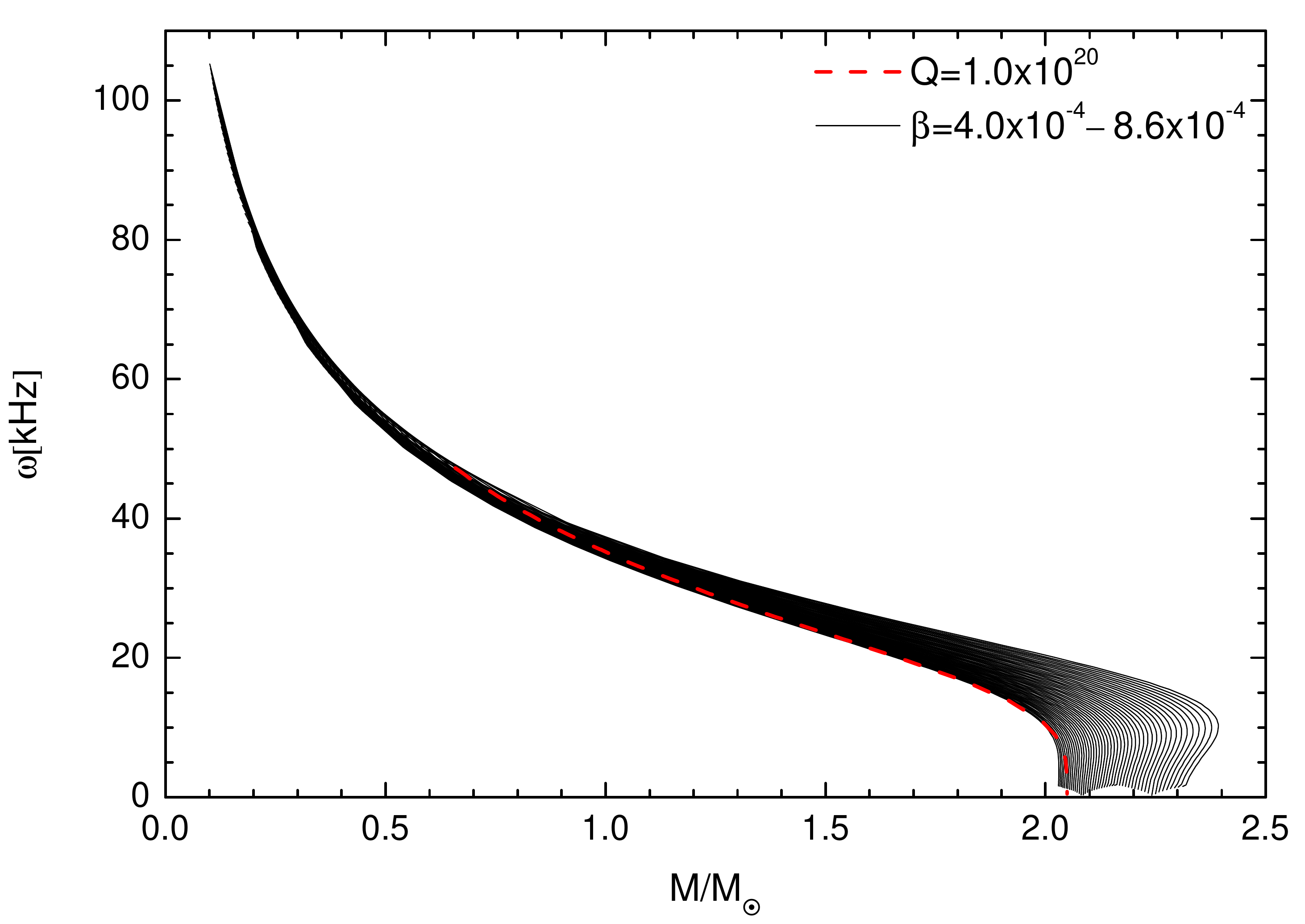}
\vspace*{-.7cm}
\caption{The top panel shows the behavior of the mass with the central energy density, whereas the one in the middle shows the variation of the total mass with the total radius, and the one at the bottom shows the eigenfrequency of the fundamental mode as a function of the total mass. As is indicated in each panel, the solid lines shown are made considering $4.0\times10^{-4}\leq\beta\leq8.6\times10^{-4}[M_{\odot}/{\rm km}^3]$ and the dashed line is constructed considering the configurations with $Q=1.0\times 10^{20}\,[\rm C]$.}
\label{rho_m_R_w}
\end{figure}

The curves used to study the hydrostatic equilibrium and stability against radial oscillations with fixed value of the total charge are built using the configurations with the same value of $Q$ found in each curve obtained in the study of hydrostatic equilibrium and stability with fixed $\beta$. This can be observed graphically in Fig.~\ref{rho_m_R_w} where each panel shows the results obtained for different values of $\beta$ between $4.0\times10^{-4}$ and $8.6\times10^{-4}[M_{\odot}/{\rm km}^3]$ (solid lines) and for a total charge $Q=1.0\times 10^{20}\,[\rm C]$ (dashed line). 

In Fig.~\ref{rho_m_R_w}, the top panel shows how the total mass changes with an increment of the central energy density. In the middle panel the mass-radius relation is shown, and at the bottom panel the behavior of the eigenfrequency of the fundamental mode with the total mass can be seen. 

\subsection{Equilibrium of charged strange quark stars with fixed total charge}\label{equi_beta_Q}

\begin{figure}[!h]
\includegraphics[scale=0.29]{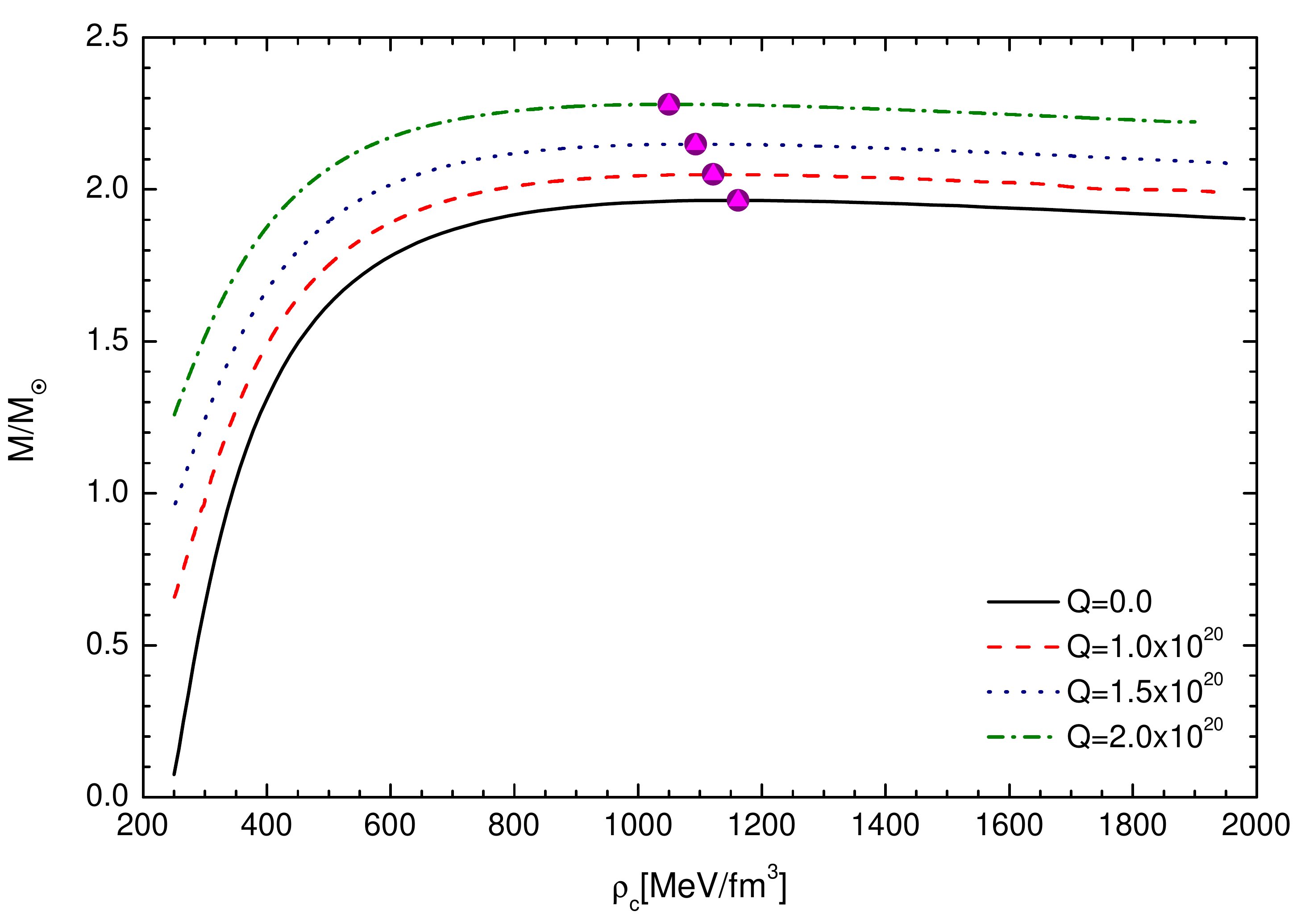}
\vspace*{-.7cm}
\caption{The total mass against central energy density for some values of the total charge. The maximum mass points and the zero eigenfrequencies are indicated respectively with full circles and full triangles. The total charge has units of $[\rm C]$.}
\label{M_rho_Q}
\end{figure}

In Fig.~\ref{M_rho_Q} the total mass as a function of the central energy density for four total charge values is plotted. The full circles indicate the maximum mass points and the full triangles represent the zero eigenfrequencies. Both geometric forms coincide at the same points. This means that these points, where $\frac{\partial M}{\partial\rho_c}\left|_{_Q}\right.=0$, separate the regions of stable and unstable stars. The configurations lying on the segments with $\frac{dM}{d\rho_c}>0$ are always stable against radial oscillations, in turn, the opposite inequality $\frac{dM}{d\rho_c}<0$ always indicates instability of charged configurations. I.e., in a system of configurations with fixed total charge, the inequalities $\frac{dM}{d\rho_c}>0$ and $\frac{dM}{d\rho_c}<0$ are necessary and sufficient conditions to determine stable and unstable configurations against radial oscillations.

\begin{figure}[!h]
\includegraphics[scale=0.29]{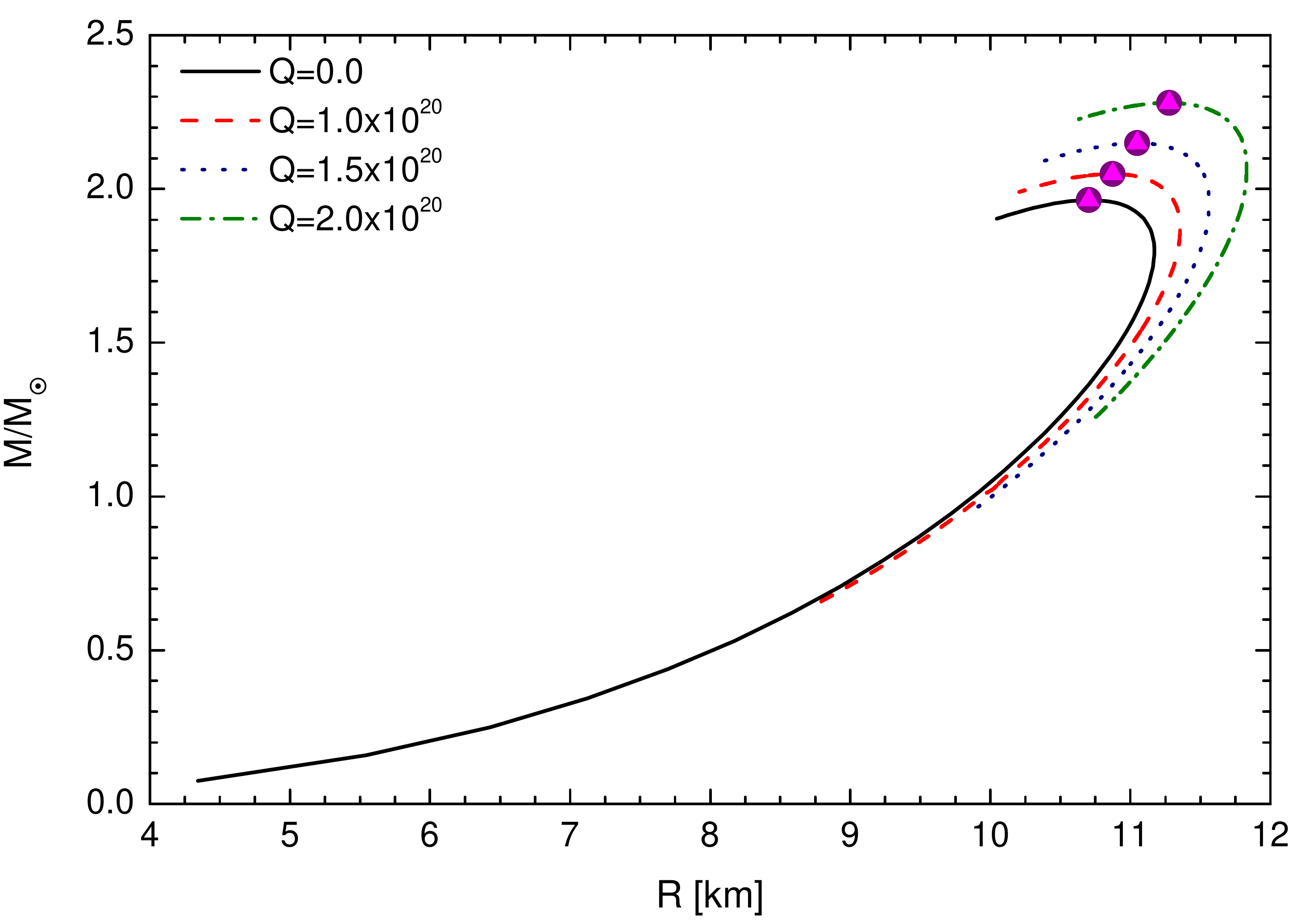}
\vspace*{-.7cm}
\caption{The total mass as a function of the total radius for some different values of the total charge. The full circles and the fill triangles indicate the maximum mass points and the zero eigenfrequencies, respectively. The total charge has units of $[\rm C]$.}
\label{R_M_Q}
\end{figure}
Fig.~\ref{R_M_Q} shows the total mass against the total radius for some different values of the total charge. The circles and triangles on the curves indicate the maximum mass points and the points where the zero eigenfrequencies are found. In this figure, as in Fig.~\ref{M_rho_Q}, the maximum mass and the zero eigenfrequency coincide at the same points.

\begin{figure}[!h]
\includegraphics[scale=0.29]{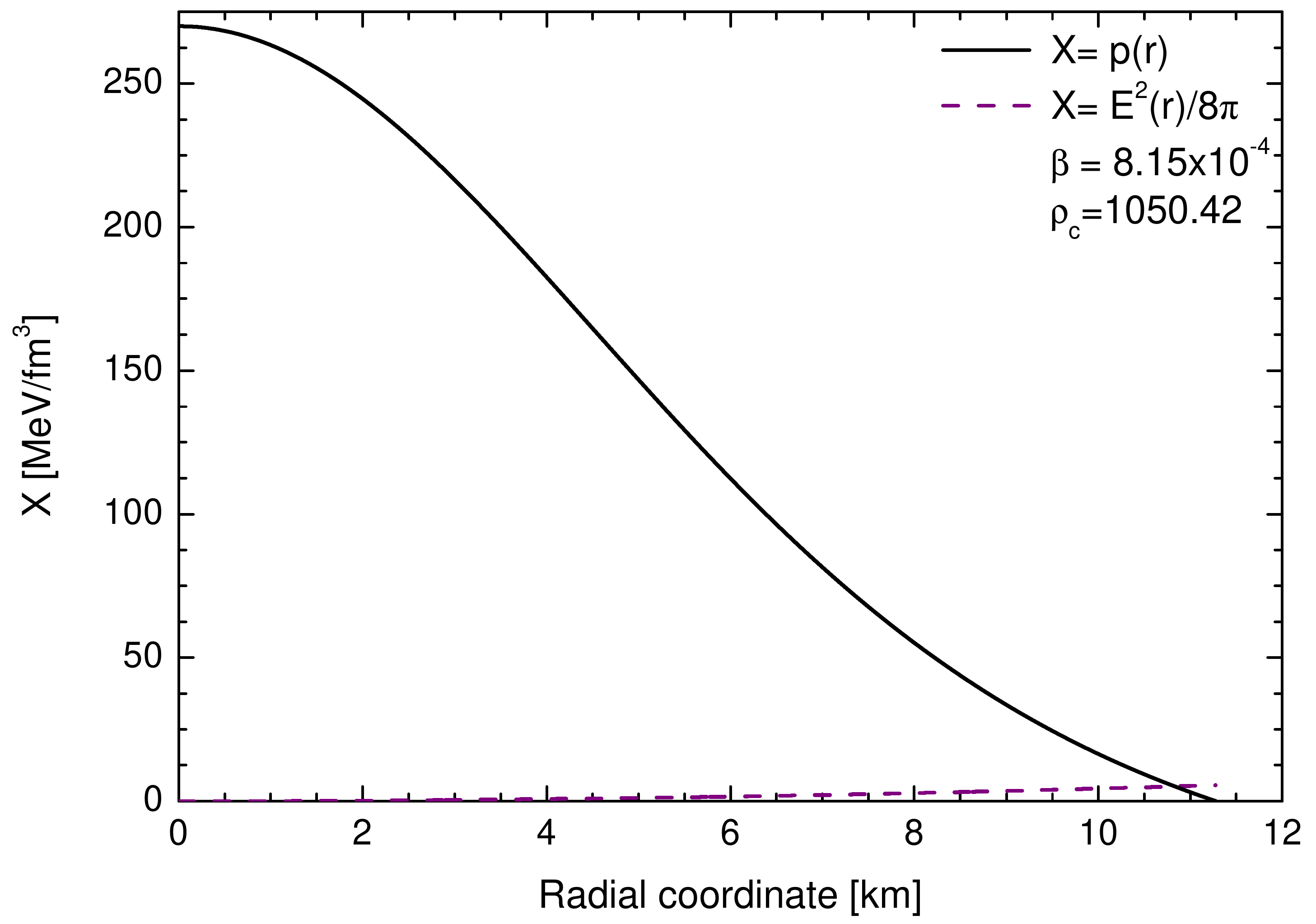}
\vspace*{-.7cm}
\caption{The radial pressure and the electrical energy density as a function of the radial coordinate in a configuration with maximum mass found in Q$=2.0\times10^{20}[\rm C]$. The units of $\beta$ and of the central energy density $\rho_c$ are $[\rm M_{\odot}/km^3]$ and $[\rm MeV/fm^3]$. }
\label{rc_X_a1}
\end{figure}

The behavior of the radial pressure and of the electrical energy density with the radial coordinate are presented in Fig.~\ref{rc_X_a1}, for a static equilibrium configuration with maximum mass found in Q=$2.0\times10^{20}[\rm C]$. The values of $\beta=8.15\times10^{-4}\,[\rm M_{\odot}/km^3]$ and $\rho_c=1050.42\,[\rm MeV/fm^3]$ are used. Note that the electrical energy density and the radial pressure are in the same order of magnitude only near the surface of the star, from this we understand that the strange quark stars are affected by the quantity of charge lying near the star's surface.

\subsection{Stability of charged strange quark stars with fixed total charge}

\begin{figure}[!h]
\includegraphics[scale=0.29]{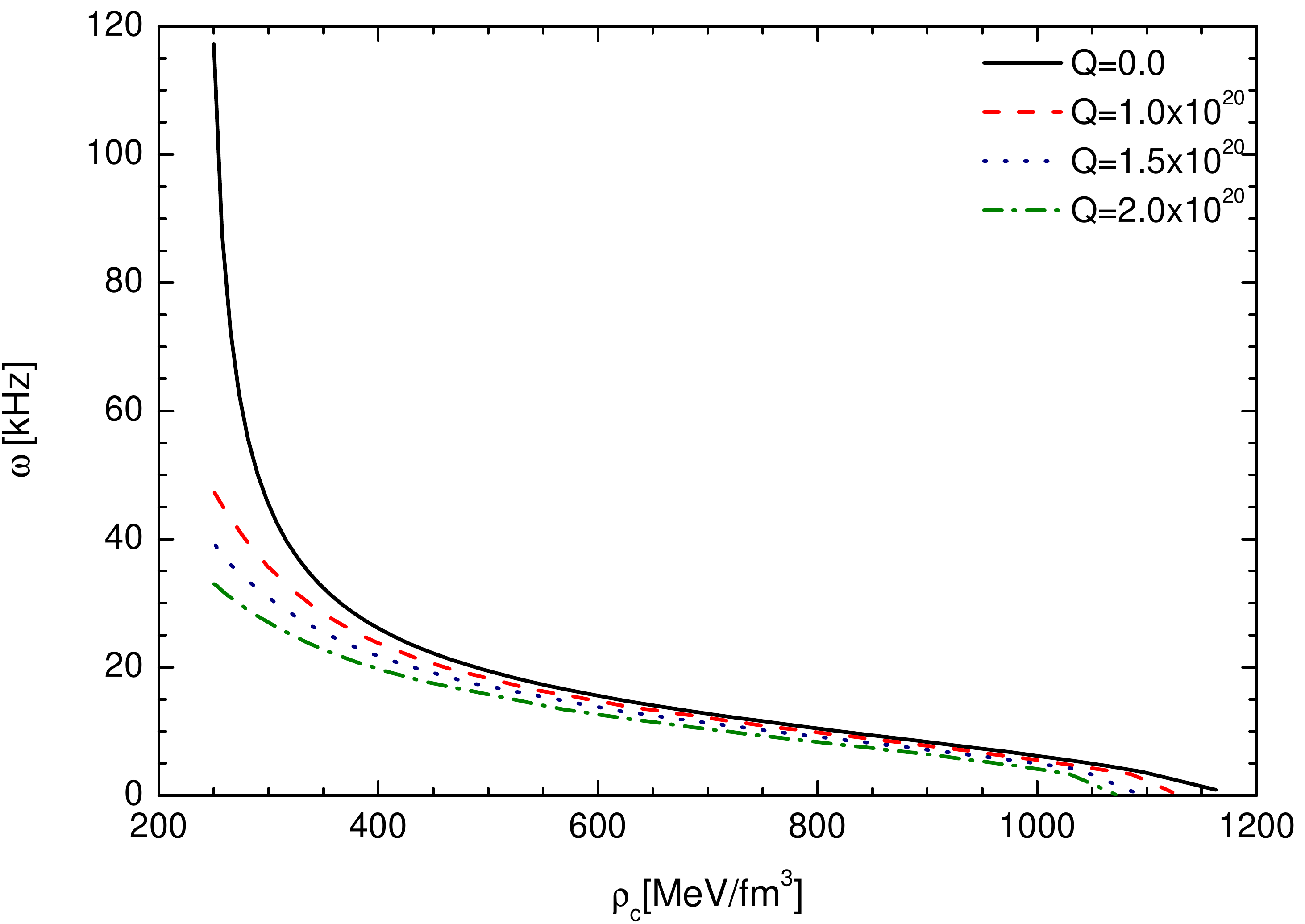}
\vspace*{-.7cm}
\caption{The eigenfrequency of the fundamental mode against the central  energy density, for some values of total charge. The total charge has units of $[\rm C]$. }
\label{w_rho_Q}
\end{figure}
The oscillation frequency as a function of the central energy density for different values of total charge is plotted in Fig.~\ref{w_rho_Q}. In the graphic we only consider values of the central energy densities where we found a positive value of $\omega$. For all values of $Q$ considered, we observe that the eigenfrequency of the fundamental mode decreases monotonically with the increment of the central energy density. This means that for larger central energy density, we have lower stability. On the other hand, the influence of the charge on the stellar stability can be also observed in Fig.~\ref{w_rho_Q}. There is a range of values of $\rho_c$ for which $\omega$ decreases with the increment of $Q$, as found in the study of hybrid stars developed in \cite{brillante2014}. In other words, for a range in central energy density the increment of electric charge helps to decrease the stability of the star.

\begin{figure}[!h]
\includegraphics[scale=0.29]{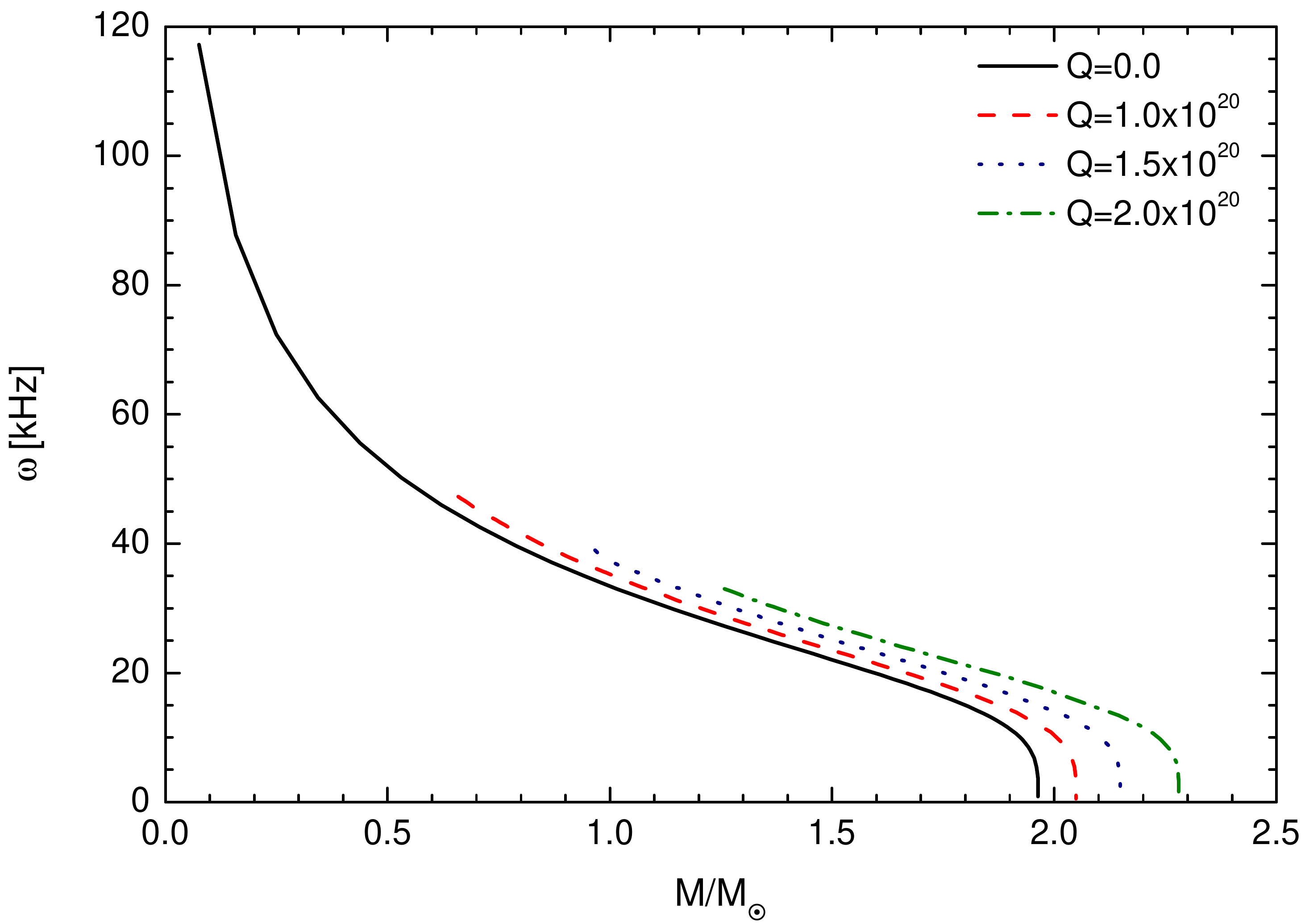}
\vspace*{-.7cm}
\caption{The eigenfrequency of the fundamental mode versus the total mass for few values of the total charge. The total charge has units of $[\rm C]$.}
\label{w_M_Q}
\end{figure}
The eigenfrequency of the fundamental mode as a function of the total mass for some values of the total charge $Q$ is seen in Fig.~\ref{omega_M} where we only consider stable configurations, i.e., charged star with $\omega$ positive. In all curves, it can be observed that the value of $\omega$ decreases monotonically with the increase of the mass $M/M_{\odot}$. Clearly we see that the zero eigenfrequency is attained for the maximum mass value. For a range of values of the mass, we have that the electric charge helps the star to become more stable against radial oscillations.

\subsection{Turning-point method for stability of charged stars}

An analysis of the stability of charged stars against radial perturbation can be done using an alternative method. The stability of charged stars can be analyzed in a manner similar to that developed in the study of the turning-point method for axisymmetric stability of uniformly rotating relativistic stars in \cite{sorkin1988}. Using the study developed in \cite{sorkin1982}, in \cite{sorkin1988} the authors found that along a sequence of rotating objects with increasing $\rho_c$ and with fixed angular momentum the equilibrium configurations with maximum mass point marks the onset of instability. Using the turning-point method and following the steps used in \cite{sorkin1988}, it must be possible to show that in a sequence of charged stars with fixed total charge and growing central energy density the point of maximum mass marks the beginning of the instability, such as it was demonstrated through the perturbation method.

\section{Dependence of the strange quark equilibrium and stability on the charge distribution}\label{supplement}

\subsection{Charge density relation}

In order to analyze the dependence on the choice of the charge distribution in our study of equilibrium and stability of strange quark stars, we consider in this section the charge density related to the energy density by the form:
\begin{equation}\label{rho_e_rho}
\rho_e=\alpha\rho,
\end{equation}
being $\alpha$ a dimensionless proportionality constant, already used in the first charged compact star self-consistent calculation solving the charged Tolman-Oppenheimer-Volkoff equation \cite{raymalheirolemoszanchin}.

The numerical method used in this study is described in Subsec.~\ref{numerical_method}. It is clear that the stellar structure equations and the radial oscillation equations are solved for different values of $\rho_c$ and $\alpha$ instead of $\rho_c$ and $\beta$ as were done above.

\subsection{Equilibrium and stability of charged strange quark stars with fixed $\alpha$}

\begin{figure}[!ht]
\centering
\includegraphics[scale=0.29]{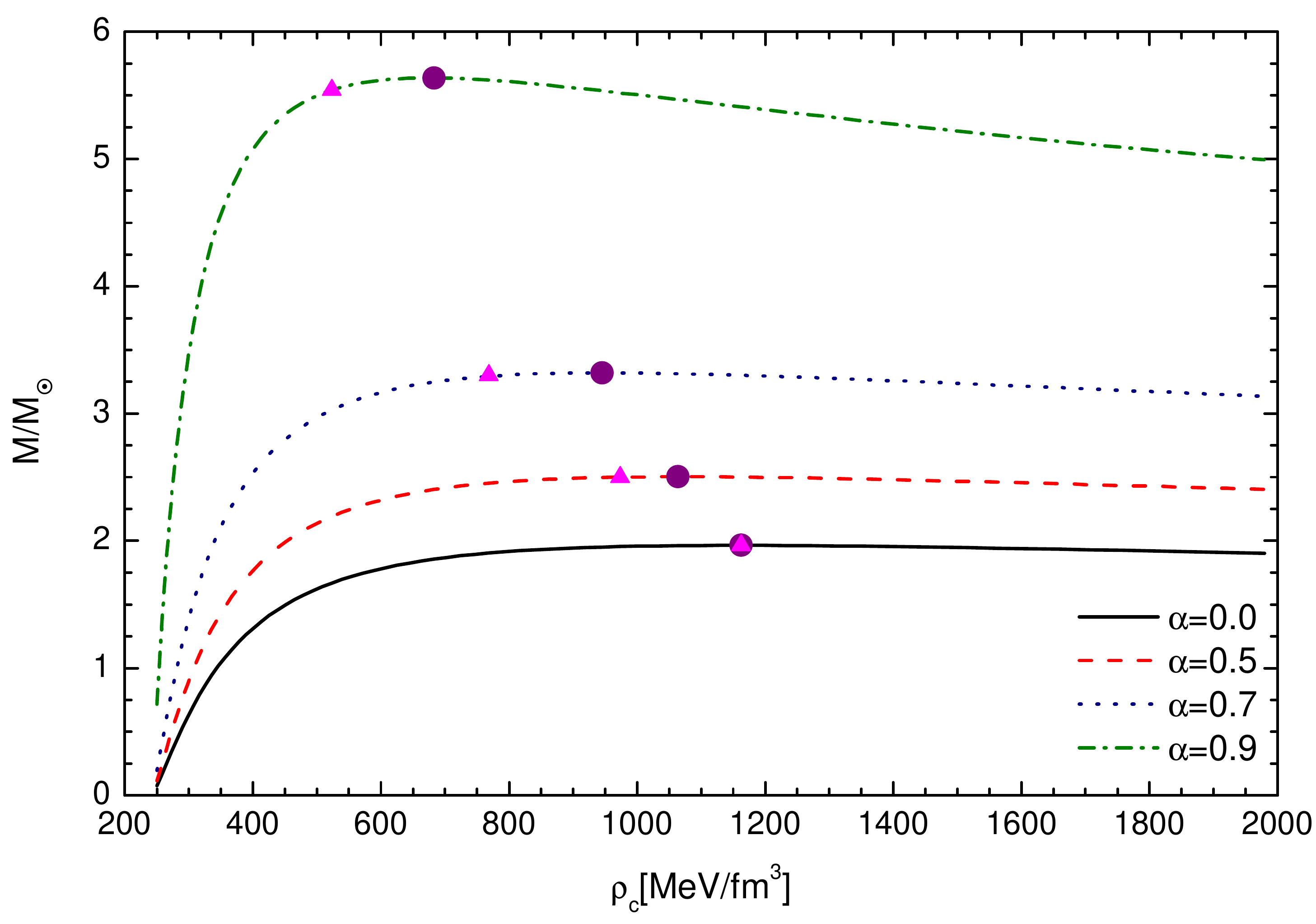}
\centering
\includegraphics[scale=0.29]{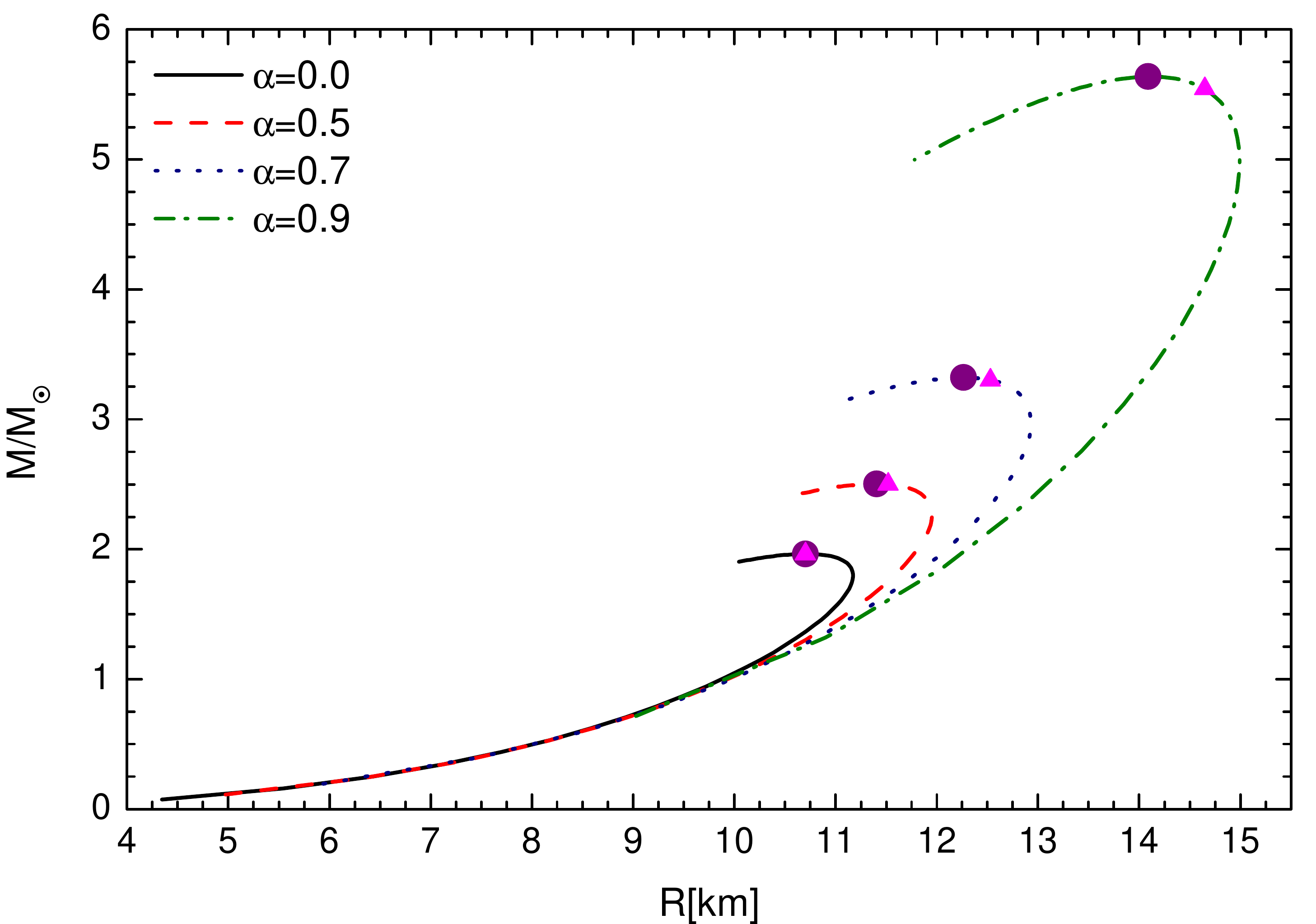}
\vspace*{-.7cm}
\caption{Top panel: the behavior of the total mass of the star in solar masses as a function of the central energy density. Bottom panel: the total mass versus the total radius of the star. In both panels, the curves are plotted considering four values of $\alpha$. The full circles represent the maximum mass points and the full triangles the places where the zero eigenfrequencies of the fundamental mode are found.}
\label{rho_m_R_alpha}
\end{figure} 

Fig.~\ref{rho_m_R_alpha} shows in the top panel the behavior of the total mass of the star as a function of the central energy density and in the bottom panel the behavior of the mass as a function of the total radius. In both panels, the curves are plotted considering four values of $\alpha$. The full circles and the full triangles over the curves indicate respectively the places where are found the maximum mass points and the zero eigenfrequency of the fundamental mode. As we can observe in the charged case ($\alpha\neq0$), the full circles and the full triangles not match at the same point indicating that the condition $\frac{dM}{d\rho_c}>0$ $\left(\frac{dM}{d\rho_c}<0\right)$ is a necessary but not a sufficient condition to determine regions made of stable (unstable) configurations.

\begin{figure}[!h]
\includegraphics[scale=0.29]{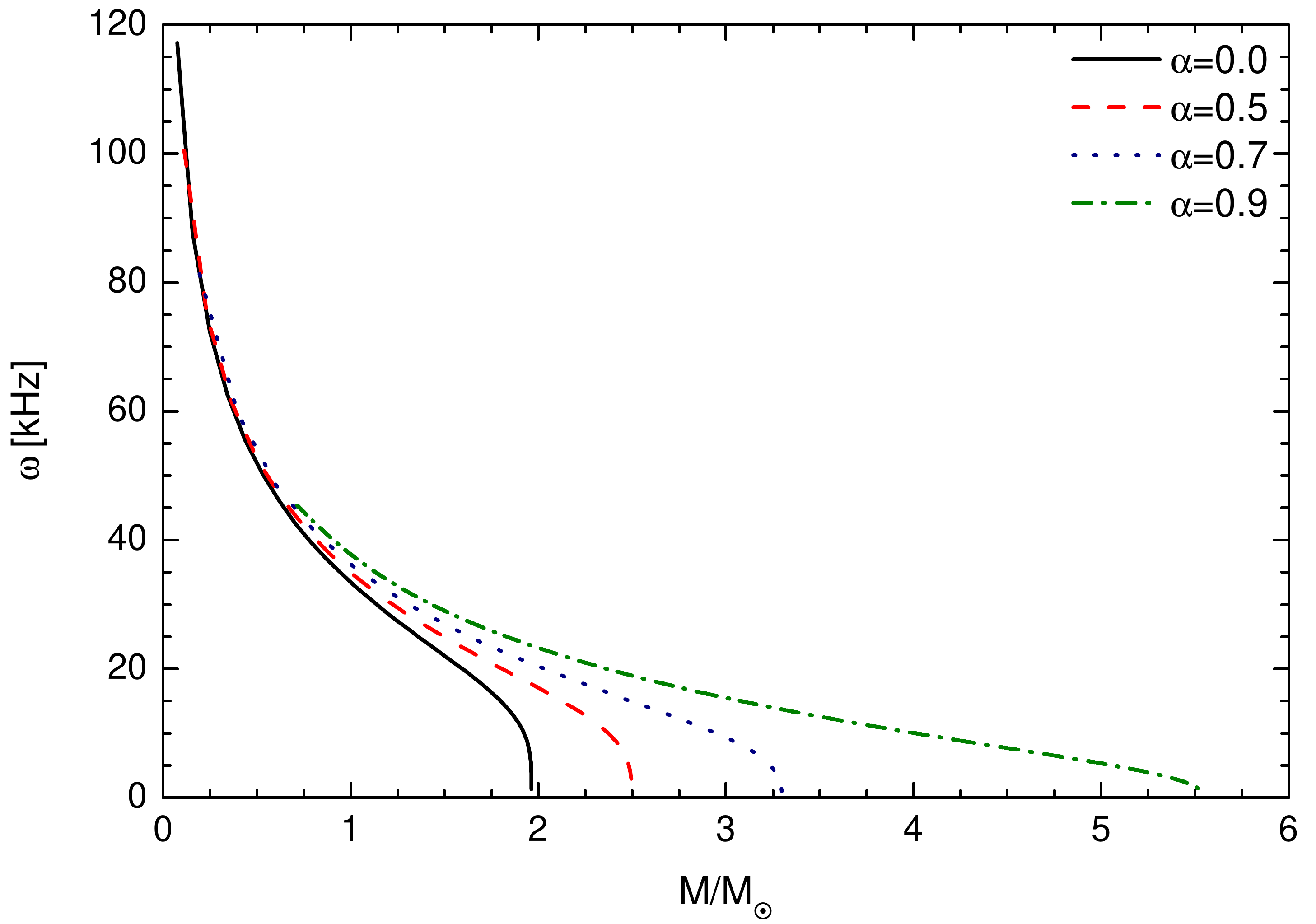}
\vspace*{-.7cm}
\caption{The eigenfrequency of the fundamental mode as a function of the stellar mass of strange charged quark stars for some values of $\alpha$. }
\label{omega_M_alpha}
\end{figure}

The eigenfrequency of the fundamental mode against the total mass is plotted in Fig.~\ref{omega_M_alpha} for four values of $\alpha$. In all cases presented the values of $\omega$ decrease monotonically with the growth of the total mass.

\subsection{Equilibrium and stability of charged strange quark stars with fixed charge}\label{equi_stabi_alpha_Q}

\begin{figure}[!h]
\centering
\includegraphics[scale=0.29]{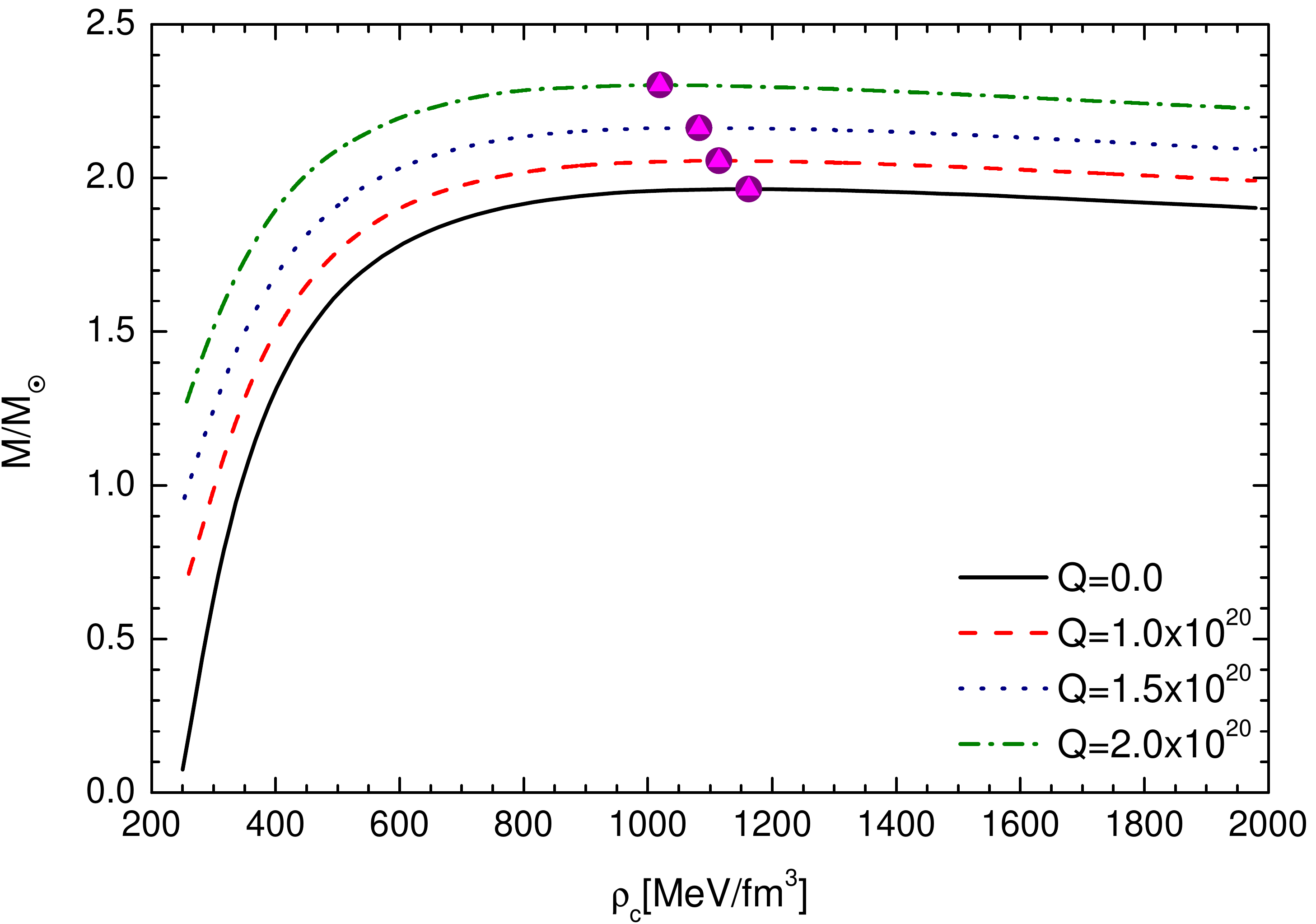}
\centering
\includegraphics[scale=0.29]{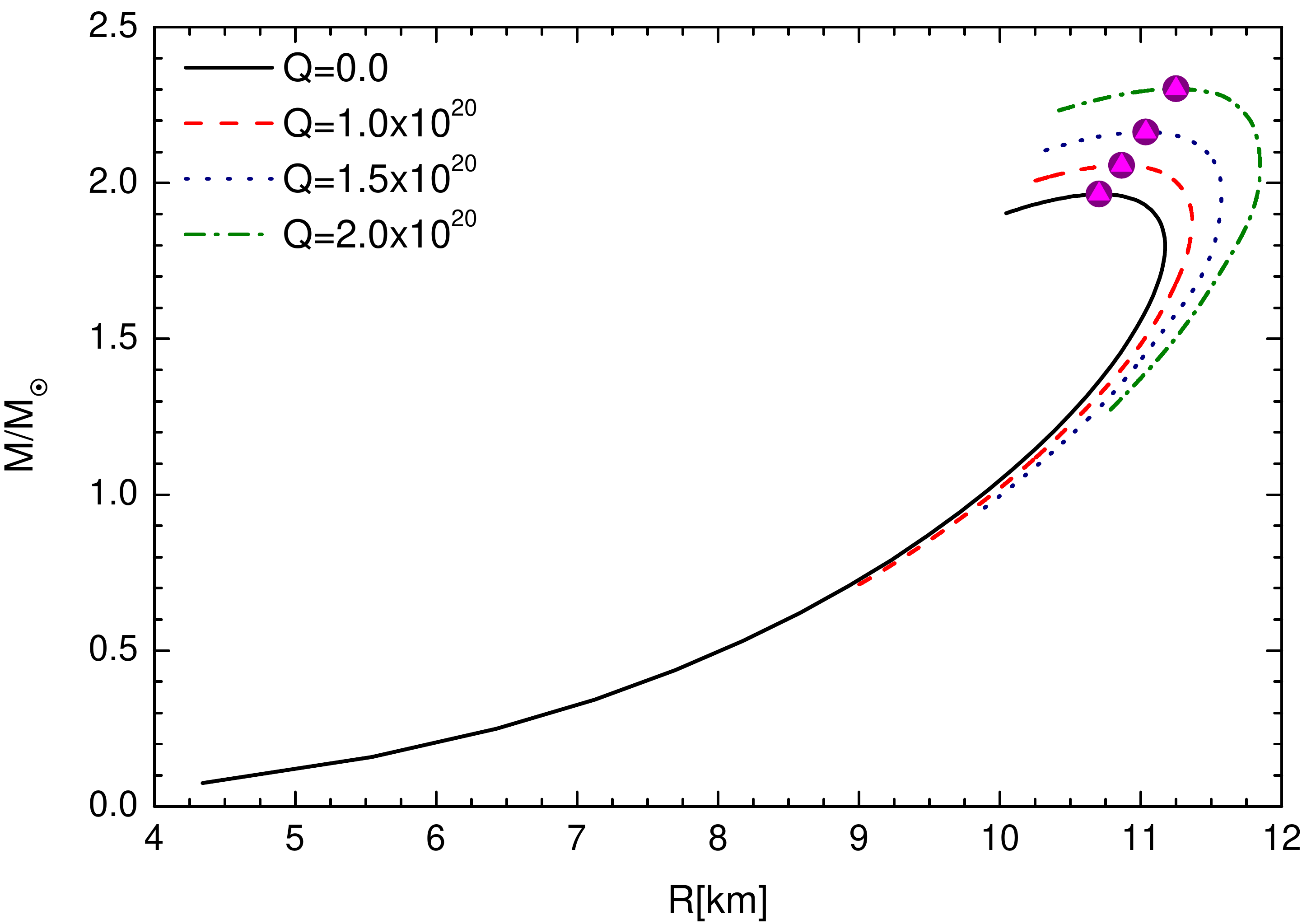}
\vspace*{-.7cm}
\caption{On the top panel and on the bottom panel are shown the total mass of the star versus of the central energy density and the total mass against the total radius of the star, respectively. As is indicated in both panels, the curves are made considering four values of total charge. The full circles represent the maximum mass points and the full triangles the places where the zero eigenfrequencies of the fundamental mode are found.}
\label{rho_m_R_alpha_Q}
\end{figure} 
The stellar mass of the charged strange quark star as a function of the central energy density and as a function of the total radius for four different values of the total charge are plotted respectively on top panel and on the bottom panel of Fig.~\ref{rho_m_R_alpha_Q}. As we can note, the maximum mass points and the zero eigenfrequency points overlap at the same place when is fixed the total charge. Analyzing the top panel, we distinguish that regions made by stable and unstable configurations against radial oscillations can be recognized through relations $\frac{dM}{d\rho_c}>0$ and $\frac{dM}{d\rho_c}<0$ when the total charge is fixed.

\begin{figure}[!h]
\includegraphics[scale=0.29]{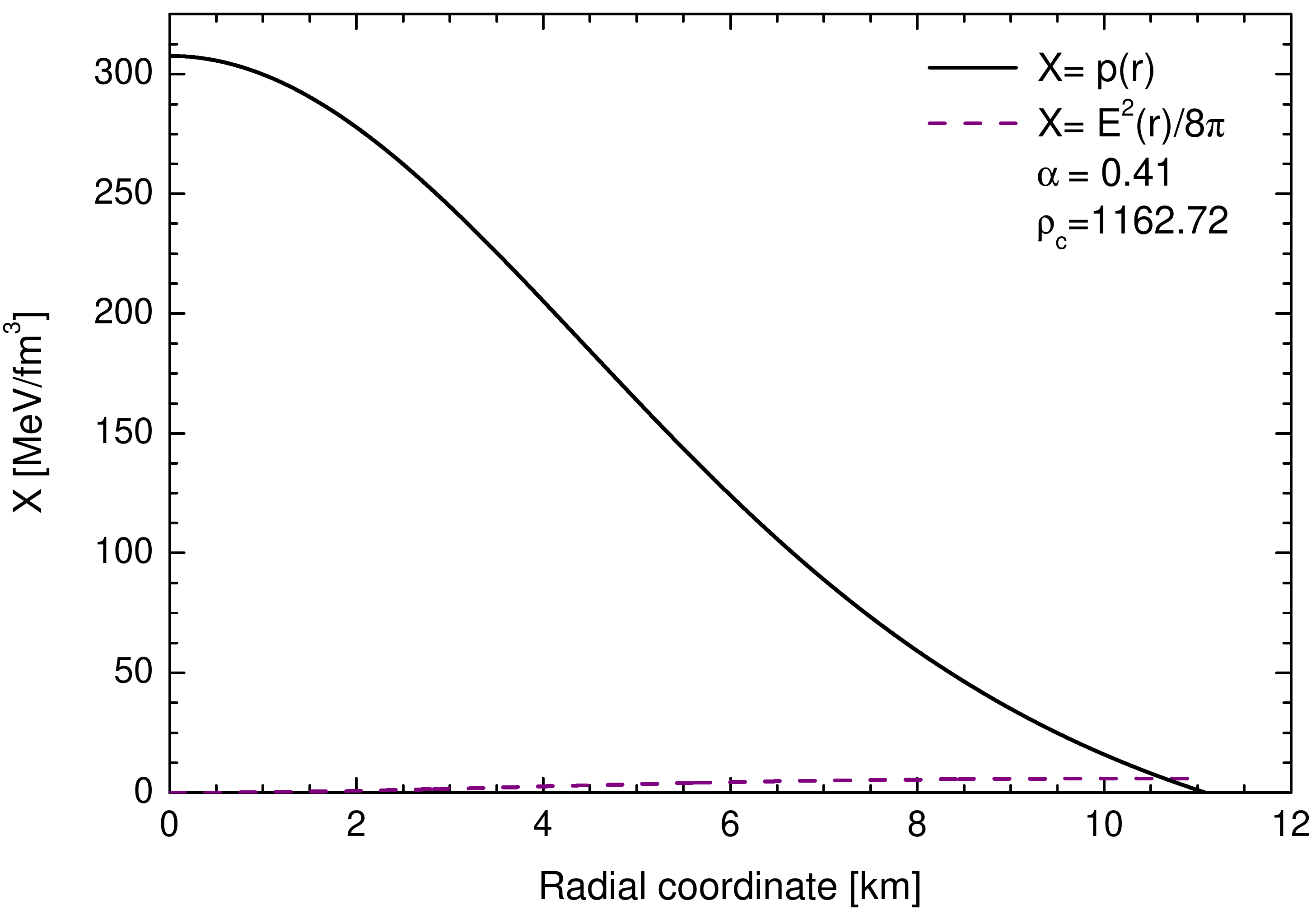}
\vspace*{-.7cm}
\caption{The radial pressure and the electrical energy density against the radial coordinate in a configuration with maximum mass found in Q$=2.0\times10^{20}[\rm C]$. The central energy density $\rho_c$ are $[\rm MeV/fm^3]$. }
\label{rc_X_alpha}
\end{figure}

From these results and the ones obtained in Subsec. \ref{equi_beta_Q} we understand that the configurations on the segments 
$\frac{dM}{d\rho_c}>0$ and $\frac{dM}{d\rho_c}<0$ are always stable and unstable against radial oscillations when the total charge is fixed, independently of the charge distribution used (either $q=\beta r^3$ or $\rho_e=\alpha\rho$). We understand this result since for a fixed charge of stellar objects, the total charge is always limited by the electrical charge distributed near the surface: as we can see observing Figs.~\ref{rc_X_a1} and \ref{rc_X_alpha} the pressure due to the electrical energy density does not vanishes, as it is the case for the radial pressure, near the surface of the spherical object. This electrical pressure needs to be balanced by gravity at star surface, and from Gauss law depends only on the total charge (not how it is distributed) and the star radius.

In fact as we already discussed, our results are similar to those found in Table I of \cite{negreiros2009} for the same total charge. In this work, the radial distribution of the electric charge is only concentrate near the star's surface.

Fig.~\ref{rc_X_alpha} shows the radial pressure and the electrical energy density as a function of the radial coordinate for an equilibrium configuration with maximum mass found in the case Q=$2.0\times10^{20}[\rm C]$. In this figure we use the values $\alpha=0.41$ and $\rho_c=1162.72\,[\rm MeV/fm^3]$. 

\begin{figure}[!h]
\includegraphics[scale=0.29]{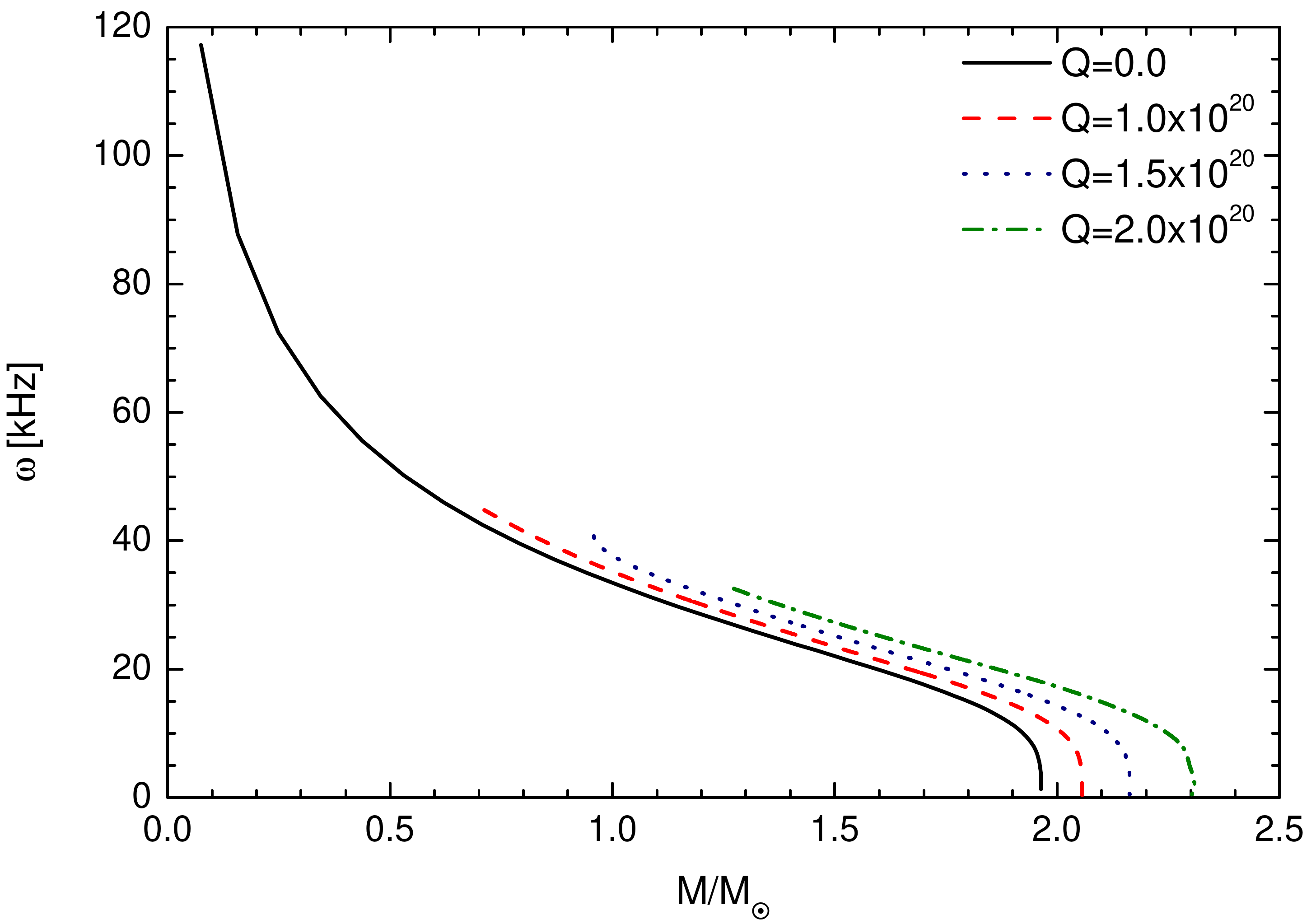}
\vspace*{-.7cm}
\caption{The eigenfrequency of the fundamental mode versus of the stellar mass of strange charged quark stars for four values of total charge. }
\label{omega_M_alpha_Q}
\end{figure}

Fig.~\ref{omega_M_alpha_Q} shows how the eigenfrequency of the fundamental mode changes with the stellar mass for some values of $\alpha$. Note in the figure that the zero eigenfrequencies are attained in the maximum mass points when is fixed the total charge.

\section{Conclusions}\label{conclusion}

In this article we study the equilibrium and stability of stars made of strange matter that follows the MIT bag model equation of state and with
an electrical charge distribution of the form $q(r)=(Q/R^3)\,r^3\equiv\beta r^3$. The con\-fi\-gu\-ra\-tions under study have spherical symmetry and are matched to the exterior Reissner-Nordstr\"om spacetime. The hydrostatic equilibrium and the stability against radial perturbation were analyzed for some different values of $\rho_c$, $\beta$ and $Q$. 

We found that the necessary total charge value to influence in the equilibrium and stability of the star is around $10^{20}[\rm C]$, the electric field produced for that total charge is around $10^{22}[\rm V/m]$. We found the electric charge that produces significant effect on the structure and stability of the object is near the star's surface, since near the surface of the star the electric energy density is not negligible compared to the radial pressure.

Using the radial perturbation method, the results indicate that for a central energy density range the stability of the star decreases with the growth of the total charge and for a total mass range the electric charge helps to increase the stability of the stars.

We also determine that the maximum mass point and the zero eigenfrequency of the fundamental mode are found in the same value of the central energy density when the total charge is fixed. This indicates that in a system of configurations with fixed total charge the stable and unstable equilibrium configurations can always be distinguished through the conditions $\frac{dM}{d\rho_c}>0$ and $\frac{dM}{d\rho_c}<0$, respectively. Finally, considering the charge density proportional to the energy density $\rho_e=\alpha\rho$, we show that the condition used to distinguish regions made of stable and unstable charged equilibrium configurations found considering the distribution of charge $q=\beta r^3$ are maintained.

\begin{acknowledgments}
\noindent 
JDVA thanks Coordena\c{c}\~ao de Aperfei\c{c}oamento de Pessoal de N\'\i vel Superior - CAPES, Brazil, for a grant.  
\end{acknowledgments}

\end{document}